\documentclass[twoside]{article}

\usepackage[latin1]{inputenc}
\usepackage{t1enc}
\usepackage{a4}
\usepackage{tabularx}
\usepackage{epsf}

\newcommand{\sig}{\:\lower0.6ex\hbox{$\stackrel{\textstyle >}{\sim}$}\:}
\newcommand{\sil}{\:\lower0.6ex\hbox{$\stackrel{\textstyle <}{\sim}$}\:}
\newcommand{\sigs}{\:\lower0.4ex\hbox{$\stackrel{\scriptstyle
      >}{\scriptstyle \sim}$}\,}
\newcommand{\sils}{\:\lower0.4ex\hbox{$\stackrel{\scriptstyle
      <}{\scriptstyle \sim}$}\,}
\def\etal{{\em et {al.}}}%

\def\bref{\vspace{4pt}\noindent\hangindent=10mm}
\def\araa{{\em ARAA}}

\def\aj{{\em AJ}}

\def\apj{{\em ApJ}}
\def\apjs{{\em ApJS}}
\def\aap{{\em A\&A}}
\def\apss{{\em ApSS}}
\def\dansssr{{\em Dokl.\ Akad.\ Nauk SSSR}}
\def\mnras{{\em MNRAS}}

\def\pta{{\em Phil.\ Trans.\ A.}}
\def\prl{{\em Phys.\ Rev.\ Lett.}}
\def\pasj{{\em PASJ}}

\def\pfl{{\em Phys.\ Fluids}}
\def\ppl{{\em Phys.\ Plasmas}}
\def\rmp{{\em Rev.\ Mod.\ Phys.}}
\def\za{{\em Z.\ Astrophys.}}

\begin{document}

\setcounter{figure}{0}
\setcounter{section}{0}
\setcounter{equation}{0}

\begin{center}

{\Large\bf

Star Formation in Turbulent Interstellar Gas }\\[0.7cm]

Ralf S.\ Klessen \\[0.17cm]

Astrophysikalisches Institut Potsdam, An der Sternwarte 16,
D-14482 Potsdam, Germany \\

e-mail: rklessen@aip.de

\end{center}

\vspace{0.5cm}

\begin{abstract}

\noindent{\it
  Understanding the star formation process is central to much of
  modern astrophysics.  For several decades it has been thought that
  stellar birth is primarily controlled by the interplay between
  gravity and magnetostatic support, modulated by ambipolar diffusion.
  Recently, however, both observational and numerical work has begun
  to suggest that supersonic interstellar turbulence rather than
  magnetic fields controls star formation.  Supersonic turbulence can
  provide support against gravitational collapse on global scales,
  while at the same time it produces localized density enhancements that
  allow for collapse on small scales. The efficiency and timescale of
  stellar birth in Galactic molecular clouds strongly depend on the
  properties of the interstellar turbulent velocity field, with slow,
  inefficient, isolated star formation being a hallmark of turbulent
  support, and fast, efficient, clustered star formation occurring in
  its absence.
}

\end{abstract}

\section{Introduction}
\label{sec:introduction}

Stars are important. They are the primary source of radiation (with
competition from the 3K black body radiation of the cosmic microwave
background and from accretion processes onto black holes in active
galactic nuclei, which themselves are likely to have formed from
stars), and of all chemical elements heavier than the H and He that
made up the primordial gas.  The Earth itself consists primarily of
these heavier elements, called metals in astronomical terminology.
Metals are produced by nuclear fusion in the interior of stars, with
the heaviest elements produced during the passage of the final
supernova shockwave through the most massive stars.  To reach the
chemical abundances observed today in our solar system, the material
had to go through many cycles of stellar birth and death.  In a
literal sense, we are star dust.

Stars are also our primary source of astronomical information and,
hence, are essential for our understanding of the universe and the
physical processes that govern its evolution. At optical wavelengths
almost all natural light we observe in the sky originates from stars.
In daytime this is more than obvious, but it is also true at night.
The Moon, the second brightest object in the sky, reflects light from
our Sun, as do the planets, while virtually every other
extraterrestrial source of visible light is a star or a collection of
stars.  Throughout the millenia, these objects have been the
observational targets of traditional astronomy, and define the
celestial landscape, the constellations.  When we look at a dark night
sky, we can also note dark patches of obscuration along the band of
the Milky Way.  These are clouds of dust and gas that block the light
from stars further away.

Since about half a century ago we know that these clouds are
associated with the birth of stars (for a historic account see Herbig
2002). The advent of new observational instruments and techniques gave
access to astronomical information at wavelengths far shorter and
longer that visible light.  It is now possible to observe astronomical
objects at wavelengths ranging high-energy $\gamma$-rays down to radio
frequencies.  Especially useful for studying these dark clouds are
radio and sub-mm wavelengths, at which they are transparent.
Observations now show that {\em all} star formation occurring in the
Milky Way is associated with these dark clouds.

These clouds are dense enough, and well enough protected from
dissociating UV radiation by self-shielding and dust scattering in
their surface layers for hydrogen to be mostly in molecular form in
their interior.  The density and velocity structure of molecular
clouds is extremely complex and follows hierarchical scaling relations
that appear to be determined by supersonic turbulent motions (e.g.\ 
Williams, Blitz, \& McKee 2000).  Molecular clouds are large, and
their masses exceed the threshold for gravitational collapse by far
when taking only thermal pressure into account.  Naively speaking,
they should be contracting rapidly and form stars at very high rate.
This is generally not observed.  The star formation efficiency of
molecular clouds in the solar neighborhood is estimated to be of order
of a few percent (e.g.\ Elmegreen 1991, McKee 1999).

For many years it was thought that support by magnetic pressure
against gravitational collapse offered the best explanation for the
low rate of star formation.  In this so called ``standard theory of
star formation'', developed by Shu (1977; and see Shu, Adams, \&
Lizano 1987), Mouschovias \& Spitzer (1976), Nakano (1976), and
others, interstellar magnetic fields prevent the collapse of gas clumps
with insufficient mass to flux ratio, leaving dense cores in
magnetohydrostatic equilibrium.  The magnetic field couples only to
electrically charged ions in the gas, though, so neutral atoms can
only be supported by the field if they collide frequently with ions.
The diffuse interstellar medium (ISM) with number densities $n$ of
order unity remains ionized highly enough so that neutral-ion
collisional coupling is very efficient (see Mouschovias 1991).  In
dense cores, where $n > 10^5$~cm$^{-3}$, ionization fractions drop
below parts per million.  Neutral-ion collisions no longer couple the
neutrals tightly to the magnetic field, so the neutrals can diffuse
through the field in a process known in astrophysics as ambipolar
diffusion.  This allows gravitational collapse to proceed in the face
of magnetostatic support, but on a timescale as much as an order of
magnitude longer than the free-fall time, drawing out the star
formation process.

Recently, however, both observational and theoretical results have
begun to cast doubt on the ``standard theory'' (for a recent
compilation see Mac~Low \& Klessen 2003). While theoretical
considerations point against singular isothermal spheres as starting
conditions of protostellar collapse as postulated by the theory (see
Whitworth \etal\ 1996, Nakano 1998, Desch \& Mouschovias 2001), there
is a series of observational findings that put other fundamental
assumptions of the ``standard theory'' into question as well. For
example, the observed magnetic field strengths in molecular cloud
cores appear too weak to provide support against gravitational collapse
(Crutcher 1999, Bourke \etal\ 2001). At the same time, the infall
motions measured around star forming cores extend too broadly (e.g.\ 
Tafalla \etal\ 1998 or Williams \etal\ 1999 for L1544), while the
central density profiles of cores are flatter than expected for
isothermal spheres (e.g.\ Bacmann \etal\ 2000).  Furthermore, the
chemically derived ages of cloud cores are comparable to the free-fall
time instead of the much longer ambipolar diffusion timescale (Bergin
\& Langer 1997). Observations of young stellar objects also appear
discordant.  Accretion rates appear to decrease rather than remain
constant, far more embedded objects have been detected in cloud cores
than predicted, and the spread of stellar ages in young clusters does
not approach the ambipolar diffusion time (as discussed in the review
by Andr{\'e} \etal\ 2000).

These inconsistencies suggest to look beyond the standard theory, and
we do so by seeking inspiration from the classical dynamical picture
of star formation which we reconsider in the light of the recent
progress in describing and understanding molecular cloud turbulence.
Rather than relying on quasistatic evolution of magnetostatically
supported objects, a new dynamical theory of star formation invokes
supersonic interstellar turbulence to control the star formation
process.  We argue that this is both sufficient to explain star
formation in Galactic molecular clouds and more consistent with
observations.

Our line of reasoning leads us first to a general introduction of the
concept of turbulence (Section \ref{sec:turbulence}), which is then
followed by an analysis of its decay properties (Section
\ref{sec:motions}). As our arguments rely to a large degree on results
from numerical models we give a brief introduction into numerical
simulations of supersonic turbulence (Section \ref{sec:self-grav}). We
then discuss how local collapse can occur in globally stable
interstellar gas clouds (Section \ref{sec:collapse}) and investigate
the physical processes that may prevent or promote this collapse
(Section \ref{sec:pplc}) leading to either more clustered or more
isolated modes of star formation (Section
\ref{sec:clustered-isolated}). We deal with the timescales of star
formation (Section \ref{sec:timescales}) and discuss how the
inclusion of magnetic fields may influence molecular cloud
fragmentation (Section \ref{sec:MHD}). We also discuss specific
predictions of the new theory of turbulent star formation for
protostellar mass accretion rates (Section \ref{sec:accretion}) and
for the resulting stellar mass spectra (Section \ref{sec:IMF}). We
then speculate about physical scales of interstellar turbulence in our
Galaxy (Section \ref{sec:scales}), and ask what sets the overall
efficiency of star formation (Section \ref{sec:eff}) and what terminates
the process on scales of individual star forming regions (Section
\ref{sec:termination}). At the end of this review (Section
\ref{sec:summary}), we summarize our results and conclude that indeed
the hypothesis that stellar birth is controlled by the complex
interplay between supersonic turbulence and self-gravity offers an
attractive pathway towards a consistent and comprehensive theory of
star formation.

\section{Turbulence}
\label{sec:turbulence}
At this point, we should briefly discuss the concept of turbulence,
and the differences between supersonic, compressible (and magnetized)
turbulence, and the more commonly studied incompressible turbulence.
We mean by turbulence, in the end, nothing more than the gas flow
resulting from random motions at many scales.  We furthermore will use
in the discussion below only the very general properties and scaling
relations of turbulent flows, focusing mainly on effects of
compressibility. For a more detailed discussion of the complex
statistical characteristics of turbulence, we refer the reader to the
book by Lesieur (1997).

Most studies of turbulence treat incompressible turbulence,
characteristic of most terrestrial applications.  Root-mean-square
(rms) velocities are subsonic, and density remains almost constant.
Dissipation of energy occurs entirely in the centers of small
vortices, where the dynamical scale $\ell$ is shorter than the length
on which viscosity acts $\ell_{\rm visc}$.  Kolmogorov (1941)
described a heuristic theory based on dimensional analysis that
captures the basic behavior of incompressible turbulence surprisingly
well, although subsequent work has refined the details substantially.
He assumed turbulence driven on a large scale $L$, forming eddies at
that scale.  These eddies interact to from slightly smaller eddies,
transferring some of their energy to the smaller scale.  The smaller
eddies in turn form even smaller ones, until energy has cascaded all
the way down to the dissipation scale $\ell_{\rm visc}$.

In order to maintain a steady state, equal amounts of energy must be
transferred from each scale in the cascade to the next, and eventually
dissipated, at a rate
\begin{equation}
\dot{E} = \eta v^3/L,
\end{equation}
where $\eta$ is a constant determined empirically. This leads to a
power-law distribution of kinetic energy $E\propto v^2 \propto
k^{-10/3}$, where $k = 2\pi/\ell$ is the wavenumber, and density does
not enter because of the assumption of incompressibility.  Most of the
energy remains near the driving scale, while energy drops off steeply
below $\ell_{\rm visc}$.  Because of the local nature of the cascade in
wavenumber space, the viscosity only determines the behavior of the
energy distribution at the bottom of the cascade below $\ell_{\rm visc}$,
while the driving only determines the behavior near the top of the
cascade at and above $L$.  The region in between is known as the
inertial range, in which energy transfers from one scale to the next
without influence from driving or viscosity.  The behavior of the flow
in the inertial range can be studied regardless of the actual scale at
which $L$ and $\ell_{\rm visc}$ lie, so long as they are well separated.
The behavior of higher order structure functions $S_p(\vec{r}) =
\langle \{v(\vec{x}) - v(\vec{x}+\vec{r})\}^p \rangle$ in
incompressible turbulence has been successfully modeled by She \&
Leveque (1994) by assuming that dissipation occurs in the filamentary
centers of vortex tubes.

Gas flows in the ISM vary from this idealized picture in a number of
important ways.  Most significantly, they are highly compressible,
with Mach numbers ${\cal M}$ ranging from order unity in the warm,
diffuse ISM, up to as high as 50 in cold, dense molecular clouds.
Furthermore, the equation of state of the gas is very soft due to
radiative cooling, so that pressure $P\propto \rho^{\gamma}$ with the
polytropic index falling in the range $0.4 < \gamma < 1.2$ (e.g.\ 
Scalo \etal\ 1998, Ballesteros-Paredes, V{\'a}zquez-Semadeni, \& Scalo
1999b, Spaans \& Silk 2000). Supersonic flows in highly compressible
gas create strong density perturbations.  Early attempts to understand
turbulence in the ISM (von Weizs\"acker 1943, 1951, Chandrasekhar
1949) were based on insights drawn from incompressible turbulence.
Although the importance of compressibility was already understood, how
to incorporate it into the theory remained unclear.  Furthermore,
compressible turbulence is only one physical process that may cause
the strong density inhomogeneities observed in the ISM. Others are
thermal phase transitions (Field, Goldsmith, \& Habing 1969, McKee \&
Ostriker 1977, Wolfire \etal\ 1995) or gravitational collapse (e.g.\ 
Wada \& Norman 1999).

In supersonic turbulence, shock waves offer additional possibilities
for dissipation.  Shock waves can transfer energy between widely
separated scales, removing the local nature of the turbulent cascade
typical of incompressible turbulence.  The spectrum may shift only
slightly, however, as the Fourier transform of a step function
representative of a perfect shock wave is $k^{-2}$, so the associated
energy spectrum should be close to $\rho v^2 \propto k^{-4}$, as was
indeed found by Porter, Pouquet, \& Woodward (1994).  However, even in
hypersonic turbulence, the shock waves do not dissipate all the
energy, as rotational motions continue to contain a substantial
fraction of the kinetic energy, which is then dissipated in small
vortices.  However, Boldyrev (2002) has proposed a theory of structure
function scaling based on the work of She \& Leveque (1994) using the
assumption that dissipation in supersonic turbulence primarily occurs
in sheet-like shocks, rather than linear filaments. This model has
been extended to describe density structure functions and exhibits in
general good agreement with numerical results (see Boldyrev, Nordlund, \&
Padoan 2002).

The driving of interstellar turbulence is neither uniform nor
homogeneous.  Controversy still reigns over the most important energy
sources at different scales, but it appears likely that isolated and
correlated supernovae dominate (Mac~Low \& Klessen 2003).  However, it
is not yet understood at what scales expanding, interacting blast
waves contribute to turbulence.  Analytic estimates have been made
based on the radii of the blast waves at late times (Norman \& Ferrara
1996), but never confirmed with numerical models (much less
experiment).

Finally, interstellar gas is magnetized.  Although magnetic field
strengths are difficult to measure, with Zeeman line splitting being
the best quantitative method, it appears that fields within an order
of magnitude of equipartition with thermal pressure and turbulent
motions are pervasive in the diffuse ISM, most likely maintained by a
dynamo driven by the motions of the interstellar gas.  A model for the
distribution of energy and the scaling behavior of strongly
magnetized, incompressible turbulence based on the interaction of
shear Alfv\'en waves is given by Goldreich \& Sridhar (1995, 1997) and
Ng \& Bhattacharjee (1996).  The scaling properties of the structure
functions of such turbulence was derived from the work of She \&
Leveque (1994) by M\"uller \& Biskamp (2000; also see Biskamp \&
M\"uller 2000) by assuming that dissipation occurs in current sheets.
A theory of very weakly compressible turbulence has been derived by
using the Mach number ${\cal M} \ll 1$ as a perturbation parameter
(Lithwick \& Goldreich 2001), but no further progress has been made
towards analytic models of strongly compressible magnetohydrodynamic
(MHD) turbulence with ${\cal M} \gg 1$.

With the above in mind, we propose that stellar birth is regulated by
interstellar turbulence and its interplay with gravity.  Turbulence,
even if strong enough to counterbalance gravity on global scales, will
usually provoke local collapse on small scales.  Supersonic turbulence
establishes a complex network of interacting shocks, where converging
flows generate regions of high density. This density enhancement can
be sufficient for gravitational instability. Collapse sets in.
However, the random flow that creates local density enhancements also
may disperse them again.  For local collapse to actually result in the
formation of stars, collapse must be sufficiently fast for the region
to `decouple' from the flow, i.e.\ it must be shorter than the typical
time interval between two successive shock passages.  The shorter this
interval, the less likely a contracting region is to survive. Hence,
the efficiency of star formation depends strongly on the properties of
the underlying turbulent velocity field, on its lengthscale and
strength relative to gravitational attraction.  This principle holds
for star formation throughout all scales considered, ranging from
small local star forming regions in the solar neighborhood up to
galaxies as a whole (see Mac~Low \& Klessen 2003).

\section{Decay and Maintenance of Supersonic Motions}
\label{sec:motions}
\begin{figure}[t]
\unitlength1.0cm
\begin{picture}(10,9.6)
\put(0.2,0.0){\epsfxsize=11.3cm \epsfbox{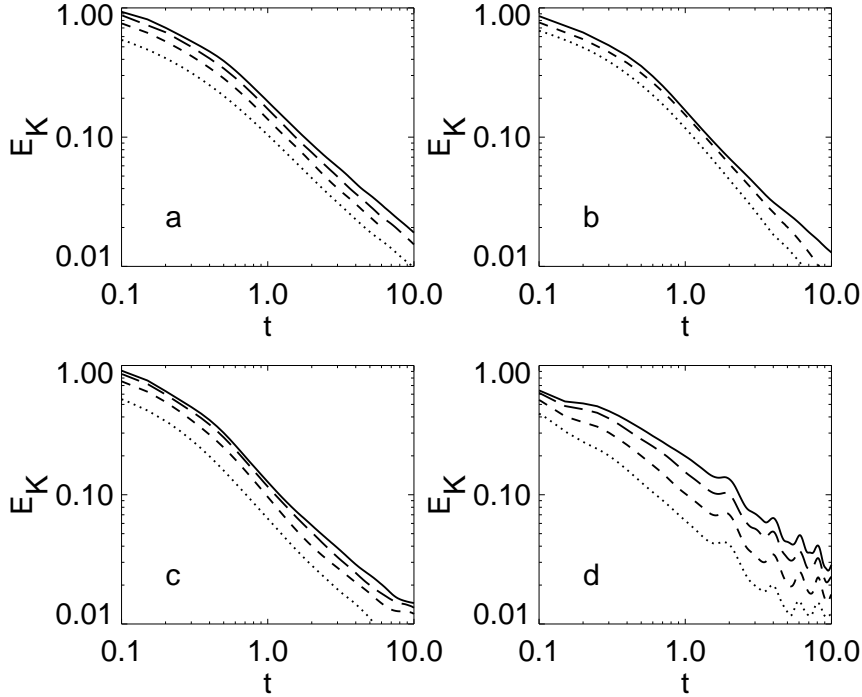}}
\end{picture}
\caption{
\label{fig:prlres} Decay of
3-dimensional supersonic turbulence for initial Mach number ${\cal
  M}=5$ and isothermal equation of state.  ZEUS models have $32^3$
({\em dotted}), $64^3$ ({\em short dashed}), $128^3$ ({\em long
  dashed}), or $256^3$ ({\em solid}) zones.  SPH models have
7000 ({\em dotted}), 50,000 ({\em short dashed}), or 350,000 ({\em
  solid}) particles. The panels show {\em a)} hydro runs with ZEUS, {\em
  b)} hydro runs with SPH, {\em c)} $A=5$ MHD runs with ZEUS, and {\em
  d)} $A=1$ MHD runs with ZEUS.  $A=v_{\rm rms}/v_{\rm A} = v_{\rm
  rms}/ (B^2/4\pi\rho)^{-1/2}$ is the
Alfv{\'e}n number. (From Mac Low \etal\ (1998)).}
\end{figure}

We first consider the question of how to maintain the observed
supersonic motions in molecular clouds.  As described above,
magnetohydrodynamic waves were generally thought to provide the means
to prevent the dissipation of interstellar turbulence.  However,
numerical models have now shown that they probably do not.
One-dimensional simulations of decaying, compressible, isothermal,
magnetized turbulence by Gammie \& Ostriker (1996) showed quick decay
of kinetic energy $K$ in the absence of driving, but found that the
quantitative decay rate depended strongly on initial and boundary
conditions because of the low dimensionality.  Mac~Low \etal\ (1998),
Stone, Ostriker \& Gammie (1998), and Padoan \& Nordlund (1999)
measured the decay rate in direct numerical simulations in three
dimensions, using a number of different numerical methods.  They
uniformly found rather faster decay, with Mac Low \etal\ (1998)
characterizing it as $K \propto t^{-\eta}$, with $0.85 < \eta < 1.1$.
A resolution and algorithm study is shown in Figure~\ref{fig:prlres}.
Magnetic fields with strengths ranging up to equipartition with the
turbulent motions (ratio of thermal to magnetic pressures as low as
$\beta = 0.025$) do indeed reduce $\eta$ to the lower end of this
range, but not below that, while unmagnetized supersonic turbulence
shows values of $\eta \approx 1 - 1.1$.

Stone \etal\ (1998) and Mac Low (1999) showed that supersonic
turbulence decays in less than a free-fall time under molecular cloud
conditions, regardless of whether it is magnetized or unmagnetized.
The hydrodynamical result agrees with the high-resolution,
transsonic, decaying models of Porter \etal\ (1994).  Mac
Low (1999) showed that the formal dissipation time $\tau_{\rm d} = K/\dot{K}$
scaled in units of the free fall time $t_{\rm ff}$ is
\begin{equation} \label{eqn:decay}
\tau_{\rm d}/\tau_{\rm ff} = \frac{1}{4 \pi \xi} \left(\frac{32}{3}\right)^{1/2}
\frac{\kappa}{{\cal M}_{\rm rms}} \simeq \,3.9 \,\frac{\kappa}{{\cal
M}_{\rm rms}},
\end{equation}
where $\xi = 0.21/\pi$ is the energy-dissipation coefficient, ${\cal
  M}_{\rm rms} = v_{\rm rms}/c_{\rm s}$ is the rms Mach number of the
turbulence, and $\kappa$ is the ratio of the driving wavelength to the
Jeans wavelength $\lambda_{\rm J}$, which is the critical scale for
gravitational collapse to set in (Jeans 1902).  In molecular clouds,
${\cal M}_{\rm rms}$ is typically observed to be of order 10 or
higher.  If the ratio $\kappa < 1$, as is probably required to
maintain gravitational support (L\'eorat \etal\ 1990), then even
strongly magnetized turbulence will decay long before the cloud
collapses and not markedly retard the collapse.

Either observed supersonic motions must be continually driven, or
molecular clouds must be less than a single free-fall time old.
Observational evidence does suggest that clouds are a few free-fall
times old, on average, though perhaps not more than two or three, so
there is likely some continuing energy input into the clouds
(Ballesteros-Paredes, Hartmann, \& V{\'a}zquez-Semadeni 1999a, Fukui
\etal\ 1999, Elmegreen 2000).

\section{Modeling Turbulence in Self-Gravitating Gas}
\label{sec:self-grav}

This leads to the question of what effects supersonic turbulence will
have on self-gravitating clouds.  Can turbulence alone delay
gravitational collapse beyond a free-fall time? Most analytical
approaches to that question are based on the assumption that the
turbulent flow is close to incompressible, and are therefore not
applicable to interstellar turbulence. However, some more recent
models have made certain progress in recovering the velocity structure
of compressible turbulence as well (Boldyrev 2002, Boldyrev, Nordlund,
\& Padoan 2002).

Numerical models of highly compressible, self-gravitating turbulence
have shown the importance of density fluctuations generated by the
turbulence to understanding support against gravity.  Early models
were done by Bonazzola \etal\ (1987), who used low resolution ($32
\times 32$ collocation points) calculations with a two-dimensional
spectral code to support their analytical results.  The hydrodynamical
studies by Passot \etal\ (1988), L{\'e}orat \etal\ (1990),
V{\'a}zquez-Semadeni, Passot, \& Pouquet (1995) and
Ballesteros-Paredes \etal\ (1999b), were also
restricted to two dimensions, and were focused on the interstellar
medium at kiloparsec scales rather than molecular clouds, although
they were performed with far higher resolution (up to $800 \times 800$
points).  Magnetic fields were introduced in these models by Passot,
V\'azquez-Semadeni, \& Pouquet (1995), and extended to three
dimensions with self-gravity (though at only $64^3$ resolution) by
V\'azquez-Semadeni, Passot, \& Pouquet (1996).  One-dimensional
computations focused on molecular clouds, including both MHD and
self-gravity, were presented by Gammie \& Ostriker (1996) and Balsara,
Crutcher \& Pouquet (2001).  Ostriker, Gammie, \& Stone (1999)
extended their work to 2.5 dimensions more recently.

These models at low resolution, low dimension, or both,
suggested several important conclusions. First, gravitational
collapse, even in the presence of magnetic fields, does not generate
sufficient turbulence to markedly slow continuing collapse. Second,
turbulent support against gravitational collapse may act at some
scales, but not others. \\ 

More recently, three-dimensional high-resolution computations by
Klessen (2000), Klessen, Heitsch, \& Mac Low (2000) and Heitsch,
Mac~Low, \& Klessen (2001a) have confirmed both of these results.
These authors used two different numerical methods: ZEUS-3D (Stone \&
Norman 1992ab), an Eulerian MHD code; and an implementation of
smoothed particle hydrodynamics (SPH; Benz 1990, Monaghan 1992), a
Lagrangian hydrodynamics method using particles as an unstructured
grid.  Both codes were used to examine the gravitational stability of
three-dimensional hydrodynamical turbulence at high resolution.  The
use of both Lagrangian and Eulerian methods to solve the equations of
self-gravitating hydrodynamics in three dimensions (3D) allowed them
to attempt to bracket reality by taking advantage of the strengths of
each approach.  This gives some protection against interpreting
numerical artifacts as physical effects (for a detailed discussion see
Klessen \etal\ 2000).

The computations discussed here were done on periodic cubes, with an
isothermal equation of state, using up to $256^3$ zones (with one
model at $512^3$ zones) or $80^3$ SPH particles. To generate turbulent
flows Gaussian velocity fluctuations are introduced with power only in
a narrow interval $k-1 \leq |\vec{k}| \leq k$, where $k =
L/\lambda_{\rm d}$ counts the number of driving wavelengths
$\lambda_{\rm d}$ in the box (Mac Low~\etal\ 1998). This offers a
simple approximation to driving by mechanisms acting on that scale.
To drive the turbulence, this fixed pattern is normalized to maintain
constant kinetic energy input rate $\dot{E}_{\rm in} = \Delta E /
\Delta t$ (Mac~Low 1999).  Self-gravity is turned on only after a
state of dynamical equilibrium has been reached.

\section{Local versus Global Collapse}
\label{sec:collapse}

First we examine the question of whether gravitational collapse can
generate enough turbulence to prevent further collapse. Hydrodynamical
SPH models initialized at rest with Gaussian density perturbations
show fast collapse, with the first collapsed objects forming in a
single free-fall time (Klessen, Burkert, \& Bate 1998; Klessen \&
Burkert 2000, 2001). Models set up with a freely decaying turbulent
velocity field behaved similarly (Klessen 2000).  Further accretion of
gas onto collapsed objects then occurs over the next free-fall time,
defining the predicted spread of stellar ages in a freely-collapsing
system.  The turbulence generated by the collapse (or virialization)
does not prevent further collapse contrary to what sometimes has been
suggested (e.g.\ by Elmegreen 1993). The presence of magnetic fields
does not change that conclusion (Balsara \etal\ 2001) as accretion
down filaments aligned with magnetic field lines onto cores can occur
readily.  This allows high mass-to-flux ratios to be maintained even
at small scales, which is necessary for supercritical collapse to
continue after fragmentation occurs.

\begin{figure}[bthp]
\unitlength1.0cm
\begin{picture}(12,11.8)
\put(1.5,0.0){\epsfxsize=9.0cm \epsfbox{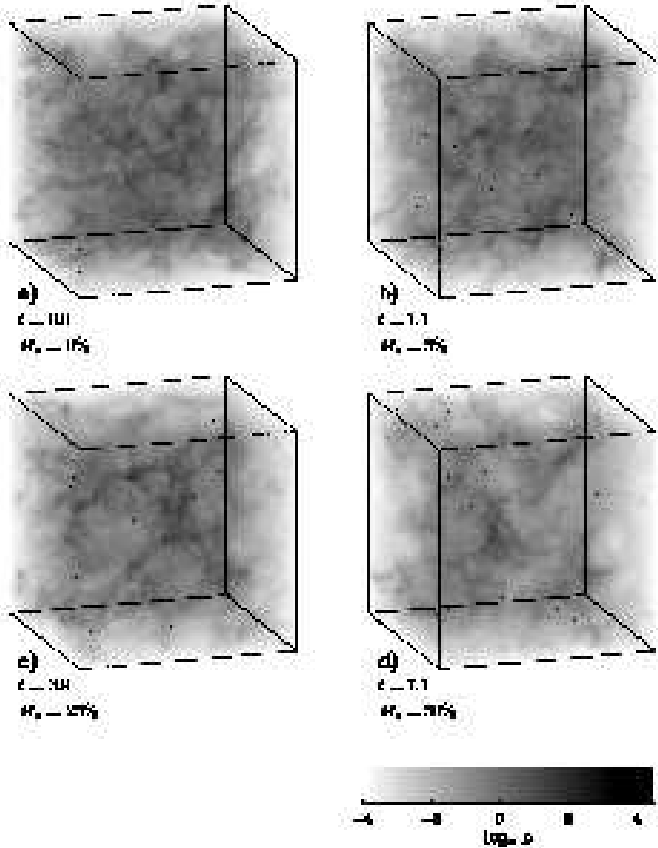}}
\end{picture}
\caption{\label{fig:3D-cubes}
  Density cubes for model ${\cal B}2h$ from Klessen \etal\ (2000),
  which is driven at intermediate wavelengths, shown (a) at the time
  when gravity is turned on, (b) when the first collapsed cores are
  formed and have accreted $M_* = 5$\% of the mass, (c) when the mass
  in dense cores is $M_* = 25$\%, and (d) when $M_* = 50$\%. Time is
  measured in units of the global system free-fall timescale
  $\tau_{\rm ff}$, dark dots indicate the location of the collapsed
  cores. }
\end{figure}

Second, we examine whether continuously driven turbulence can provide
support against gravitational collapse.  The models of driven,
self-gravitating turbulence by Klessen \etal\ (2000) and Heitsch
\etal\ (2001a) show that {\em local} collapse occurs even when the
turbulent velocity field carries enough energy to counterbalance
gravitational contraction on global scales.  An example of local
collapse in a globally supported cloud is given in
Figure~\ref{fig:3D-cubes}. A hallmark of global turbulent support is
isolated, inefficient, local collapse.

Thus, highly compressible turbulence does both, it promotes as well as
prevents collapse.  Its net effect is to inhibit collapse globally,
while at the same time promoting it locally. The resolution to this
apparent paradox lies in the requirement that any substantial
turbulent support must come from supersonic flows, as otherwise
pressure support would be at least equally important.  Supersonic
flows compress the gas in shocks. In isothermal gas with density
$\rho$ the postshock gas has density $\rho' = {\cal M}^2 \rho$, where
${\cal M}$ is the Mach number of the shock.  The turbulent Jeans
length $\lambda_{\rm J} \propto \rho'^{-1/2}$ in these density
enhancements, so it drops by a factor of ${\cal M}$ in isothermal
shocks making shock compressed gas clumps more susceptible to
gravitational collapse. On the other hand, if we consider the system
on scales exceeding the lengthscale of turbulence (i.e.\ in the limit
of microturbulence), we can follow the classical picture that treats
turbulence as an additional pressure and define an effective sound
speed $c_{{\rm s},e\!\!\!\;f\!\!f}^2 = c_{\rm s}^2 + v_{\rm rms}^2/3$
(Chandrasekhar 1949). The critical mass for gravitational collapse,
the Jeans mass $M_{\rm J} \propto \rho^{-1/2} c_{\rm s}^3$, then
strongly increases with the turbulent rms velocity dispersion $v_{\rm
  rms}$, so that for $v_{\rm rms} \gg c_{\rm s}$ turbulence ultimately
does inhibit collapse on global scales. Between these two scales,
there is a broad intermediate region, especially for long wavelength
driving, where local collapse can occur despite global support.

Klessen \etal\ (2000) demonstrated that turbulent support can
completely prevent collapse only when it can support not just the
average density, but also these high-density shocked regions, a point
that was appreciated already by Elmegreen (1993) and
V\'azquez-Semadeni \etal\ (1995).  Two criteria must be fulfilled: the
rms velocity must be sufficiently high for the turbulent Jeans
criterion to be met in these regions, and the driving wavelength
$\lambda_{\rm d} < \lambda_{\rm J}(\rho')$.  If these two criteria are
not fulfilled, the high-density regions collapse, although the
surrounding flow remains turbulently supported.  The efficiency of
collapse depends on the properties of the supporting turbulence.
Sufficiently strong driving on short enough scales can prevent local
collapse for arbitrarily long periods of time, but such strong driving
may be rather difficult to arrange in a real molecular cloud.
Furthermore, if we assume that stellar driving sources have an
effective wavelength close to their separation, then the condition
that driving acts on scales smaller then the Jeans wavelength in
`typical' shock generated gas clumps requires the presence of an
extraordinarily large number of stars evenly distributed throughout
the cloud, with typical separation 0.1 pc in Taurus, or only 350 AU in
Orion.  This is not observed. Very small driving scales seem also to
be at odds with the observed large-scale velocity fields at least in
some molecular clouds (e.g.\ Ossenkopf \& Mac~Low 2002).

\section{Promotion and Prevention of Local Collapse}
\label{sec:pplc}

\begin{figure}[t]
\unitlength1.0cm
\begin{picture}(16,7.0)
\put(1.0,0.0){\epsfxsize=9cm \epsfbox{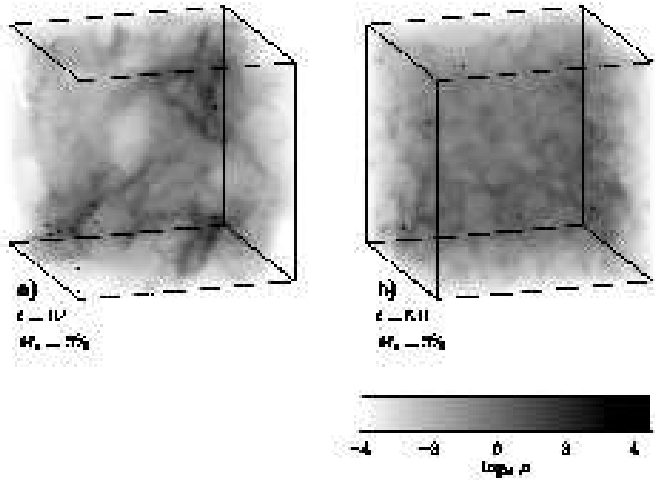}}
\end{picture}
\caption{\label{fig:3D-cubes-1..2+7..8} Density cubes for (a) a model
of large-scale driven turbulence (${\cal B}1h$) and (b) a model of
small-scale driven turbulence (${\cal B}3$) at dynamical stages where
the core mass fraction is $M_* = 5$\%. Compare with
Figure~\ref{fig:3D-cubes}b. Together they show the influence of
different driving wavelengths for otherwise identical physical
parameters. Larger-scale driving results in collapsed cores in more
organized structure, while smaller-scale driving results in more
randomly distributed cores. Note the different times at which $M_*
= 5\%$ is reached. (From Klessen \etal\ 2000.)}
\end{figure}

The origin of local collapse can also be understood in terms of a
timescale argument. Roughly speaking, the lifetime of a clump is
determined by the interval between two successive passing shocks: the
first creates it, while if the second is strong enough, it disrupts
the clump again if it has not already collapsed (Klein, McKee \&
Colella 1994, Mac Low \etal\ 1994).  If its lifetime is long enough, a
Jeans unstable clump can contract to sufficiently high densities to
effectively decouple from the ambient gas flow. It then becomes able
to survive the encounter with further shock fronts (e.g.\ Krebs \&
Hillebrandt 1983), and continues to accrete from the surrounding gas,
forming a dense core.  The weaker the passing shocks, and the greater
the separation between them, the more likely that collapse will occur.
Equivalently, weak driving and long typical driving wavelengths
enhance collapse.  The influence of the driving wavelength is more
pronounced, however, because individual shocks sweep up more mass when
the typical wavelength is longer, so density enhancements resulting
from the interaction of shocked layers will have larger masses, and so
are more likely to exceed their local Jeans limit.  Turbulent driving
mechanisms that act on large scales will produce large coherent
structures (filaments of compressed gas with embedded dense cores) on
relatively short timescales compared to small-scale driving even if
the total kinetic energy in the system is the same. Examples of the
density structure of long and small-wavelength driving, respectively,
are given in Figure \ref{fig:3D-cubes-1..2+7..8}, which can be directly
compared to Figure \ref{fig:3D-cubes}b.

\begin{figure}[th]
\unitlength1.0cm
\begin{picture}(16,7.3)
\put(-3.5,-11.0){\epsfxsize=21cm \epsfbox{ 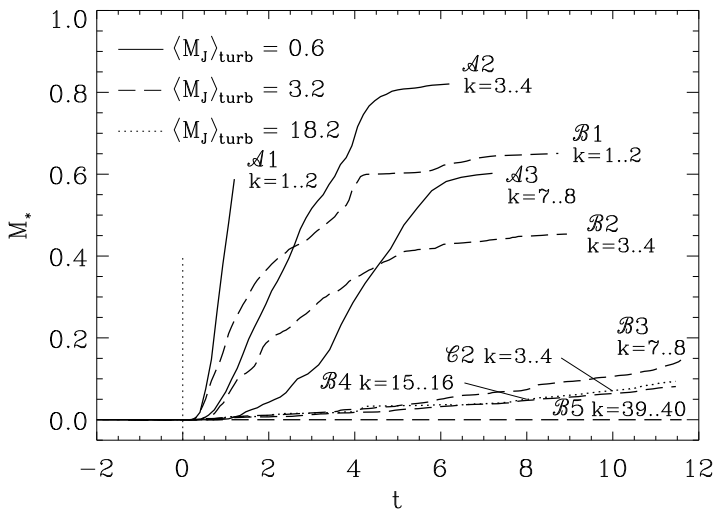}}
\end{picture}
\caption{\label{fig:accretion-history} Fraction $M_*$ of mass accreted
in dense cores as function of time for different models of
self-gravitating supersonic turbulence. The models differ by driving
strength and driving wavenumber, as indicated in the figure. The mass
in the box is initially unity, so the solid curves are formally
unsupported, while the others are formally supported.  The figure
shows how the efficiency of local collapse depends on the scale and
strength of turbulent driving.  Time is measured in units of the
global system free-fall timescale $\tau_{\rm ff}$. Only a model
driven strongly at scales smaller than the Jeans wavelength
$\lambda_J$ in shock-compressed regions shows no collapse at
all. (From Klessen \etal\ 2000. }
\end{figure}

Further insight of how local collapse proceeds comes from examining
the mass growth rates in each model.  Figure\ 
\ref{fig:accretion-history} shows the accretion history for three sets
of models from Klessen \etal\ (2000). The driving strength increases
from ${\cal A}$ over ${\cal B}$ to ${\cal C}$, but is held constant
for each set of models with the effective driving wavelength
$\lambda_{\rm d}$ being varied.  All models show local collapse,
except at the extreme end, when $\lambda_{\rm d} < \lambda_{\rm
  J}(\rho')$ (model ${\cal B}5$).

The cessation of strong accretion onto cores occurs long before all
gas has been accreted.  This is because the time that dense cores
spend in shock-compressed, high-density regions decreases with
increasing driving wavenumber and increasing driving strength.  In the
case of long wavelength driving, cores form coherently in high-density
regions associated with one or two large shock fronts that can
accumulate a considerable fraction of the total mass of the system.
The overall accretion rate is high and cores spend sufficient time in
this environment to accrete a large fraction of the total mass in the
region.  Any further mass growth has to occur from chance encounters
with other dense regions.  In the case of short wavelength driving,
the network of shocks is tightly knit. Cores form in shock generated
clumps of small masses because individual shocks are not able to sweep
up much matter. Furthermore, in this rapidly changing environment the
time interval between the formation of clumps and their destruction is
short.  The period during which individual cores are located in
high-density regions where they are able to accrete at high rate is
short as well. So altogether, the global accretion rates are small and
saturate at lower values of $M_*$ as the driving wavelength is
decreased.

\section{Clustered versus Isolated Star Formation}
\label{sec:clustered-isolated}

Different star formation regions present different distributions of
protostars and pre-main sequence stars.  In some regions, such as the
Taurus molecular cloud, stars form isolated from other stars,
scattered throughout the cloud (Mizuno \etal\ 1995).  In other
regions, they form in clusters, as in L1630 in Orion (Lada 1992), or
even more extremely in starburst regions such as 30 Doradus (Walborn
\etal\ 1999; for a review see Zinnecker \etal\ 1993).

Numerical simulations of self-gravitating turbulent clouds demonstrate
that the {\em length scale} and {\em strength} at which energy is
inserted into the system determine the structure of the turbulent flow
and therefore the locations at which stars are most likely to form.
Large-scale driving leads to large coherent shock structures (see
e.g.\ Figure\ \ref{fig:3D-cubes-1..2+7..8}a). Local collapse occurs
predominantly in filaments and layers of shocked gas and is very
efficient in converting gas into stars. This leads to what we can
identify as `clustered' mode of star formation: stars form in coherent
aggregates and clusters. Even more so, this applies to regions of
molecular gas that have become decoupled from energy input. As
turbulence decays, these regions begin to contract and form dense
clusters of stars with very high efficiency on about a free-fall time
scale (Klessen \etal\ 1998, Klessen \& Burkert 2000).  The same holds
for insufficient support, i.e.~for regions where energy input is not
strong enough to completely balance gravity. They too will contract to
form dense stellar clusters.

The `isolated' mode of star formation occurs in molecular cloud regions that
are supported by driving sources that act on {\em small} scales and in an
incoherent or stochastic manner. In this case, individual shock induced
density fluctuations form at random locations and evolve more or less
independently of each other. The resulting stellar population is widely
dispersed throughout the cloud and, as collapsing clumps are 
exposed to frequent shock interaction, the overall star formation rate
is low.

\begin{figure}[th]
\unitlength1.0cm
\begin{picture}(16,8.5)
\put( -0.3,-5.7){\epsfxsize=10.9cm \epsfbox{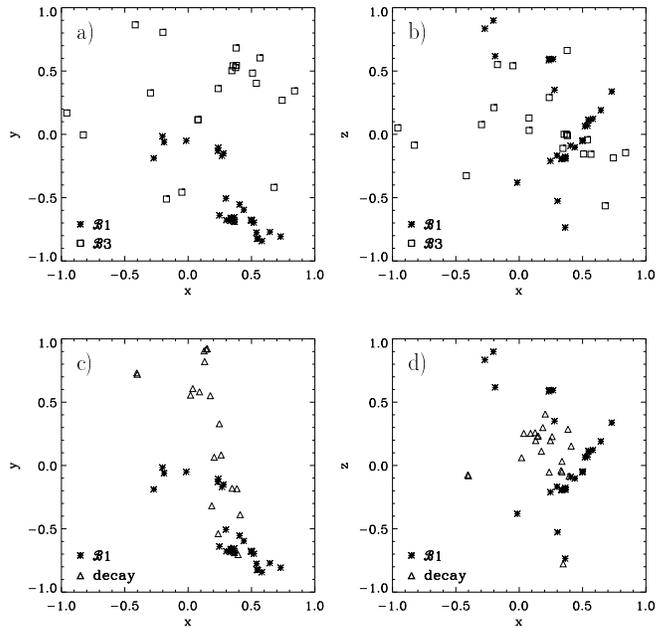}}
\end{picture}
\caption{\label{fig:2D-projection} Comparison of collapsed core
locations between two globally stable models with different driving
wavelength projected into (a) the $xy$-plane and into (b) the
$xz$-plane.  ${\cal B}1$ with $k=1-2$ is driven at large scales, and
${\cal B}3$ with $k=7-8$ is driven at small ones. Plots (c) and (d)
show the core locations for model ${\cal B}1$ now contrasted with a
simulation of decaying turbulence from Klessen (2000).  The snapshots
are selected such that the mass accumulated in dense cores is $M_*
\sil 20$\%. Note the different times needed for the different models
to reach this point.  For model ${\cal B}1$ data are taken at $t=1.1$,
for ${\cal B}3$ at $t=12.3$. The simulation of freely decaying
turbulence is shown at $t=1.1$. All times are normalized to the global
free-fall timescale of the system. (From Klessen \etal\ 2000.) }
\end{figure}

These points are illustrated in Figure~\ref{fig:2D-projection},
which shows the distribution of collapsed cores in several models with
strong enough turbulence to formally support against collapse.
Coherent, efficient local collapse occurs in model ${\cal B}1$, where
the turbulence is driven strongly at long wavelengths
(compare with Figure~\ref{fig:3D-cubes-1..2+7..8}).
Incoherent, inefficient collapse occurs in model ${\cal B}3$, on the
other hand, where turbulence is driven at small scales.
Individual cores form independently of
each other at random locations and random times,
are widely distributed throughout the entire volume, and
exhibit considerable age spread.
In the decaying turbulence model, once the kinetic energy level has
decreased sufficiently, all spatial modes of the system contract
gravitationally, including the global ones (Klessen 2000). As in the
case of large-scale shock compression, stars form more or less
coevally in a limited volume with high efficiency.

\begin{figure}[th]
\unitlength1.0cm
\begin{picture}(16,4.3)
\put( -0.3,-5.8){\epsfxsize=10.9cm \epsfbox{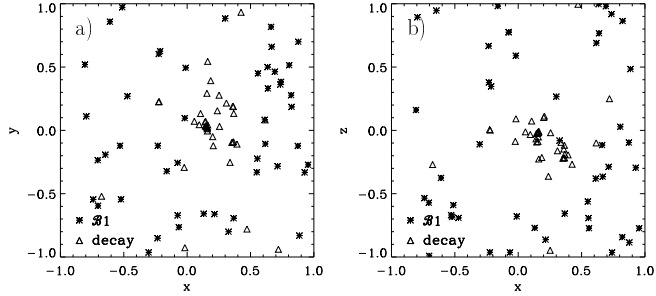}}
\end{picture}
\caption{\label{fig:2D-projection-later-times} Core positions for
  model ${\cal B}1$ ($k=1-2$) and the decay model when the core mass
  fraction is $M_* \approx 65$\%, projected into (a) the $xy$-plane
  and (b) the $xz$-plane (compare with Figure~\ref{fig:2D-projection}c
  \& d).  For ${\cal B}1$ the time is $t=8.7$ and for decay model
  $t=2.1$. Whereas the cluster in ${\cal B}1$ is completely dissolved
  and the stars are widely dispersed throughout the computational
  volume, the cluster in the decay simulation remains bound. (From
  Klessen \etal\ 2000.)}
\end{figure}

Despite the fact that both turbulence driven on large scales and freely
decaying turbulence lead to star formation in aggregates and clusters,
Figure~\ref{fig:2D-projection-later-times} suggests a possible way to
distinguish between them.
Decaying turbulence typically leads to the formation of a bound
stellar cluster, while aggregates associated with large-scale,
coherent, shock fronts often have higher velocity dispersions that
result in their complete dispersal. 
Note, however, that at the late stages of dynamical evolution shown in
Figure~\ref{fig:2D-projection-later-times}, the model becomes less
appropriate, as feedback from newly formed stars is not
included. Ionization and outflows from the stars formed first will
likely retard or even prevent the accretion of the remaining gas onto
protostars, possibly preventing a bound cluster from forming even in
the case of freely decaying turbulence.

The control of star formation by supersonic turbulence gives rise to a
continuous but articulated picture. There may not be physically distinct
modes of star formation, but qualitatively different behaviors do appear
over the range of possible turbulent flows. The apparent
dichotomy between a clustered mode of star formation and an
isolated one, as discussed by Lada (1992) for L1630 and  Strom, Strom, \&
Merrill (1993) for L1641, disappears, if a different balance between
turbulent strength and gravity holds at the relevant length scales in
these different clouds.

Turbulent flows tend to have hierarchical structure (Lesieur 1997)
which may explain the hierarchical distribution of stars in star
forming regions shown by statistical studies of the distribution of
neighboring stars in young stellar clusters (e.g.\ in Taurus, see Larson
1995).
Hierarchical clustering seems to be a common feature of all star
forming regions (e.g.\ Efremov \& Elmegreen 1998). It is a natural
outcome of turbulent fragmentation.

\section{Timescales of Star Formation}
\label{sec:timescales}
\begin{figure}[thp]
\unitlength1.0cm
\begin{picture}(16,9.5)
\put(0.0,0.0){\epsfxsize=11cm \epsfbox{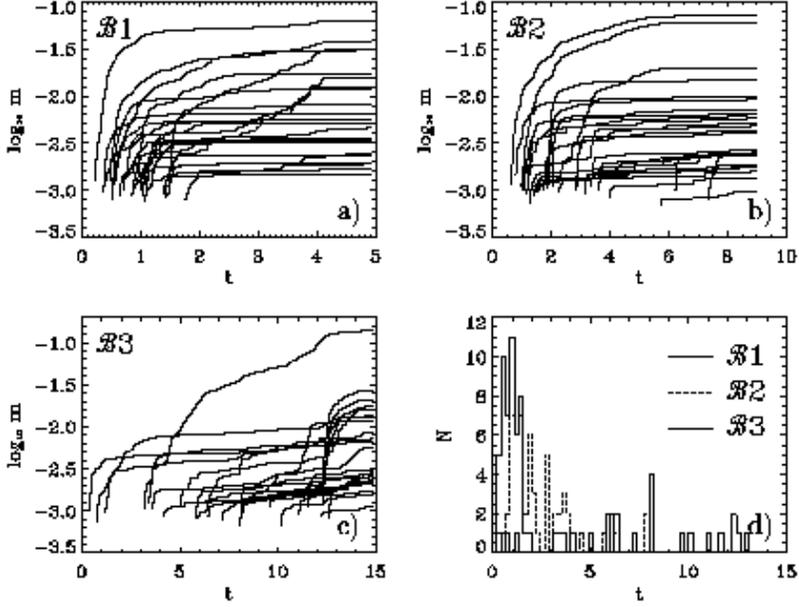}}
\end{picture}
\caption{\label{fig:core-formation-histogram} 
  Masses of individual protostars as function of time in SPH models
  (a) ${\cal B}1$ driven at large scales with $k=1-2$ driving, (b)
  ${\cal B}2$ with $k=3-4$ driving, i.e.\ at intermediate scales, and
  (c) ${\cal B}3$ with $k=7-8$ small-scale driving.  The curves
  represent the formation and accretion histories of individual
  protostars. For the sake of clarity, only every other core is shown
  in (a) and (b), whereas in (c) the evolution of every single core is
  plotted.  Time is given in units of the global free-fall time
  $\tau_{\rm ff}$. Note the different timescale in each plot. In the
  depicted time interval models ${\cal B}1$ and ${\cal B}2$ reach a
  core mass fraction $M_* =70$\%, and both form roughly 50 cores.
  Model ${\cal B}3$ reaches $M_* =35$\% and forms only 25 cores.
  Figure (d) compares the distributions of formation times. The age
  spread increases with decreasing driving scale showing that
  clustered core formation should lead to a coeval stellar population,
  whereas a distributed stellar population should exhibit considerable
  age spread. (From Klessen \etal\ 2000.)  }
\end{figure}
Turbulent control of star formation predicts that stellar clusters
form predominantly in regions that are insufficiently supported by
turbulence or where only large-scale driving is active.  In the
absence of driving, molecular cloud turbulence decays more quickly
than the free-fall timescale $\tau_{\rm ff}$ (Eq.~\ref{eqn:decay}),
so dense stellar clusters will form on the free-fall timescale. 
Even in the presence of support from large-scale driving, substantial
collapse still occurs within a few free-fall timescales, see Figure 
and~\ref{fig:core-formation-histogram}a.  If the dense cores followed
in these models continue to collapse on a short timescale to build up
stellar objects in their centers, then this directly implies the star
formation timescale.  Therefore the age distribution will be roughly
$\tau_{\rm ff}$ for stellar clusters that form coherently with high
star formation efficiency.  When scaled to low densities, say $n({\rm
H}_2) \approx 10^2\,{\rm cm}^{-3}$ and $T\approx10\,$K, the global
free-fall timescale in the models is $\tau_{\rm ff} = 3.3 \times
10^6\,$years.  If star forming clouds such as Taurus indeed have ages
of order $\tau_{\rm ff}$, as suggested by Ballesteros-Paredes \etal\
(1999), then the long star formation time  computed here is quite
consistent with the very low star formation efficiencies seen in
Taurus (e.g.\ Leisawitz \etal\ 1989, Palla \& Stahler 2000, Hartmann
2001), as the cloud simply has not had time to form many stars.  In
the case of high-density regions, $n({\rm H}_2) \approx 10^5\,{\rm
cm}^{-3}$ and $T\approx10\,$K, the dynamical evolution proceeds much
faster and the corresponding free-fall times drops to $\tau_{\rm
ff} = 10^5\,$years.  These values are indeed supported by
observational data such as the formation timescale of the Orion
Trapezium cluster.  It is inferred to stem from gas of density $n({\rm
H}_2) \sil 10^5\,{\rm cm}^{-3}$, and is estimated to be less than
$10^6$ years old (Hillenbrand \& Hartmann 1998).  The age spread in
the models increases with increasing driving wavenumber $k$ and
increasing $\langle M_{\rm J} \rangle _{\rm turb}$, as shown in
Figure~\ref{fig:core-formation-histogram}. Long periods of core
formation for globally supported clouds appear consistent with the low
efficiencies of star-formation in regions of isolated star formation,
such as Taurus, even if they are rather young objects with ages of
order $\tau_{\rm ff}$.

\section{Effects of Magnetic Fields}
\label{sec:MHD}

So far, we concentrated on the effects of purely hydrodynamic
turbulence. How does the picture discussed here change, if we consider
the presence of magnetic fields?  Magnetic fields may alter the
dynamical state of a molecular cloud sufficiently to prevent
gravitationally unstable regions from collapsing (McKee 1999).  They
have been hypothesized to support molecular clouds either
magnetostatically or dynamically through MHD waves.

\begin{figure}[th]
\unitlength1.0cm
\begin{picture}(16,7.6)
\put( 0.5, 0.0){\epsfxsize=11cm \epsfbox{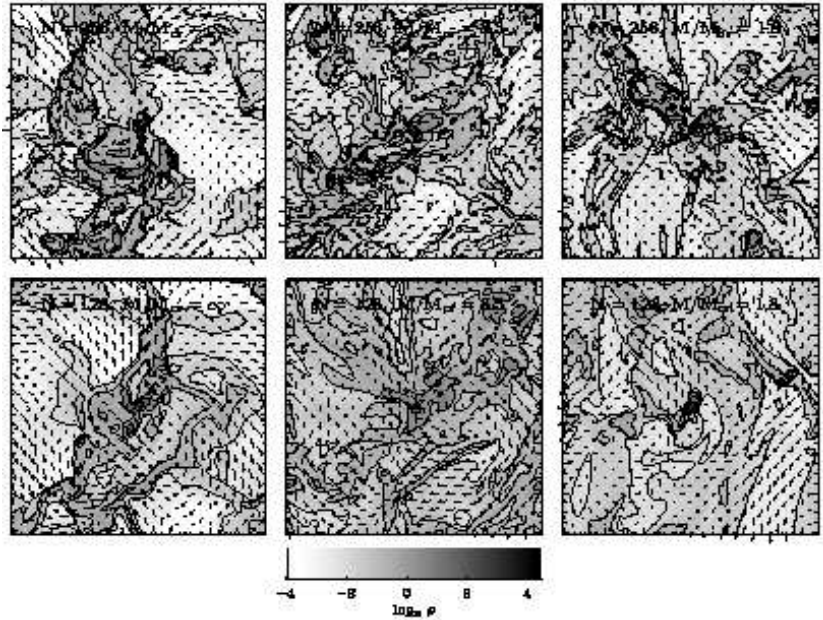}}
\end{picture}
\caption{\label{fig:2Dslices}Two-dimensional slices of $256^3$ models
from Heitsch \etal\ (2001a) driven at large scales with wavenumbers
$k=1-2$ hard enough that the mass in the box represents only 1/15
$\langle M_{\rm J}\rangle_{\rm turb}$, and with initially vertical
magnetic fields strong enough to give critical mass fractions as
shown.  The slices are taken at the location of the zone with the
highest density at the time when $10$\% of the total mass has been
accreted onto dense cores. The plot is centered on this zone.  Arrows
denote velocities in the plane. The length of the largest arrows
corresponds to a velocity of $v \sim 20 c_s$. The density greyscale is
given in the colorbar. As fields become stronger, they influence the
flow more, producing anisotropic structure. (From Heitsch \etal\ 2001a.)
}
\end{figure}

\begin{figure}[th]
\unitlength1.0cm
\begin{picture}(16, 8.0)
\put( 1.5,-0.3){\epsfxsize=8cm \epsfbox{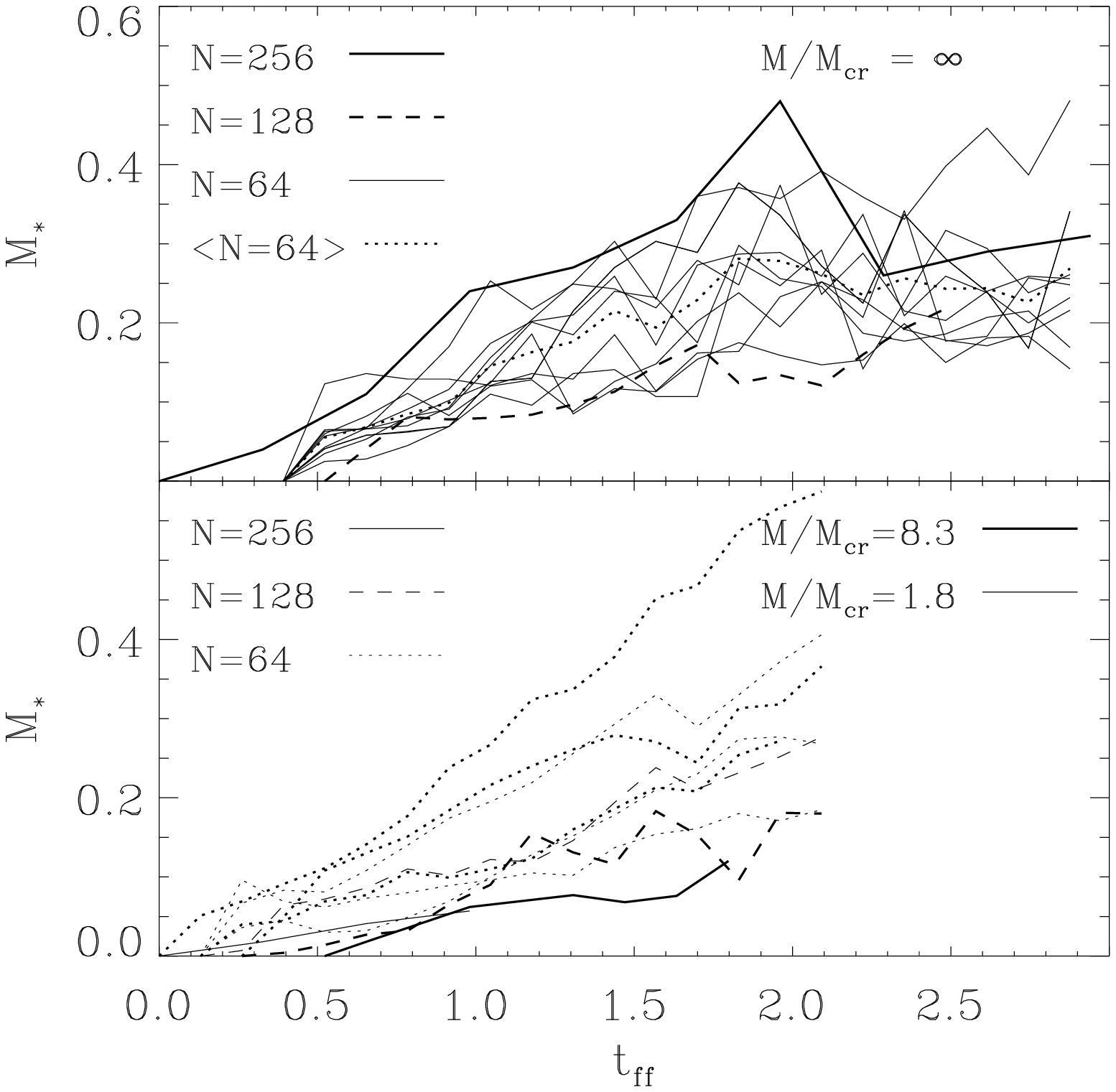}}
\end{picture}
\caption{\label{fig:variance}
  {\em Upper panel:} Core-mass accretion rates for $10$ different
  low-resolution models ($N=64^3$ cells) of purely hydrodynamic
  turbulence with equal parameter set but different realizations of
  the turbulent velocity field. The thick line shows a ``mean
  accretion rate'', calculated from averaging over the sample. For
  comparison, higher-resolution runs with identical parameters but
  $N=128^3$ and $N=256^3$ are shown as well.  The latter one can be
  regarded as an envelope for the low resolution models.  {\em Lower
    panel:} Mass accretion rates for various models with different
  magnetic field strength and resolution.  Common to all models is the
  occurrence of local collapse and star formation regardless of the
  detailed choice of parameters, as long as the system is
  magnetostatically supercritically (Heitsch
  \etal\ 2001a).}
\end{figure}

Mouschovias \& Spitzer (1976) derived an expression for the
critical mass-to-flux ratio in the center of a cloud for magnetostatic
support. Assuming ideal MHD, a self-gravitating cloud of
mass $M$ permeated by a uniform flux
$\Phi$ is stable if the mass-to-flux ratio
\begin{equation}
  \frac{M}{\Phi} <
  \left(\frac{M}{\Phi}\right)_{\rm cr}
  \label{eqn:magstatsup}
\end{equation}
with $(M/\Phi)_{\rm cr}= c_{\Phi} G^{-1/2}$. The exact value depends
on the geometry and the field and density distribution of the cloud.
A cloud is termed {\em subcritical} if it is magnetostatically stable
and {\em supercritical} if it is not.  Mouschovias \& Spitzer (1976)
determined that $c_\Phi = 0.13$ for spherical clouds. Assuming a
constant mass-to-flux ratio in a cylindrical region results in $c_\Phi
= 1/(2\pi) \approx 0.16$ (Nakano \& Nakamura 1978). Without any other
mechanism of support such as turbulence acting along the field lines,
a magnetostatically supported cloud will collapse to a sheet which
then will be supported against further collapse. Fiege \& Pudritz
(2000) discussed a sophisticated version of this magnetostatic
support mechanism, in which poloidal and toroidal fields aligned in
the right configuration could prevent a cloud filament from
fragmenting and collapsing.

Investigation of the second alternative, support by MHD waves,
concentrates mostly on the effect of Alfv\'{e}n waves, as they (1) are
not as subject to damping as magnetosonic waves and (2) can exert a
force along the mean field, as shown by Dewar (1970) and Shu
\etal\ (1987). This is because Alfv\'{e}n waves are transverse
waves, so they cause perturbations $\delta \vec{B}$ perpendicular to
the mean magnetic field $\vec{B}$. McKee \& Zweibel (1995) argue that
Alfv\'{e}n waves can even lead to an isotropic pressure, assuming that
the waves are neither damped nor driven. However, in order to support
a region against self-gravity, the waves would have to propagate
outwardly, rather than inwardly, which would only further compress the
cloud. Thus, as Shu \etal\ (1987) comment, this mechanism requires a
negative radial gradient in wave sources in the cloud.

It can be demonstrated (e.g.\ Heitsch \etal\ 2001a) that supersonic
turbulence does not cause a magnetostatically supported region to
collapse, and vice versa, that in the absence of magnetostatic
support, MHD waves cannot completely prevent collapse, although they
can retard it to some degree.  In the case of a subcritical region
with $M < M_{cr}$ sheets of high density gas form perpendicular to the
field lines.  Turbulence can shift the sheets along the field lines
without changing the mass-to-flux ratio, but collapse does not occur,
because the shock waves cannot sweep gas across field lines and the
entire region is initially supported magnetostatically.

A supercritical cloud with $M > M_{cr}$ could only be stabilized by
MHD wave pressure.  This is insufficient to completely prevent
gravitational collapse, as shown in Figure \ref{fig:2Dslices}. The
effect of the magnetic field on the morphology of the cloud is week,
and collapse occurs in all models of unmagnetized and magnetized
turbulence regardless of the numerical resolution and magnetic field
strength as long as the system is magnetically supercritical. This is
shown more quantitatively in Figure~\ref{fig:variance}. Effects of
numerical resolution make themselves felt in different ways in
hydrodynamical and MHD models.  In the hydrodynamical case, higher
resolution results in thinner shocks and thus higher peak densities.
These higher density peaks form cores with deeper potential wells that
accrete more mass and are more stable against disruption.  Higher
resolution in the MHD models, on the other hand, better resolves
short-wavelength MHD waves, which apparently can delay collapse, but
not prevent it.  This result extends to models with $512^3$ zones
(Heitsch \etal\ 2001b).

\section{Mass Growth of Protostellar Cores}
\label{sec:accretion}

Supersonic turbulence is able to produce star forming regions that
vary enormously in scale. The most likely outcome of turbulent
molecular cloud fragmentation in the Milky Way are stellar aggregates
or clusters (Adams \& Myers 2001).  The number density of protostars
and protostellar cores in the extreme cases the can be high enough for
mutual dynamical interaction to become important.  This has important
consequences for the mass growth history of individual stars and the
subsequent dynamical evolution of the nascent stellar cluster, because
this introduces a further degree of stochasticity to the star
formation process in addition to the statistical chaos associated with
turbulence and turbulent fragmentation in the first place.
  
\begin{figure}[tbh]
\unitlength1cm
\begin{picture}(11.0,6.5)
\put(  0.0,  0.0){\epsfxsize=12cm \epsfbox{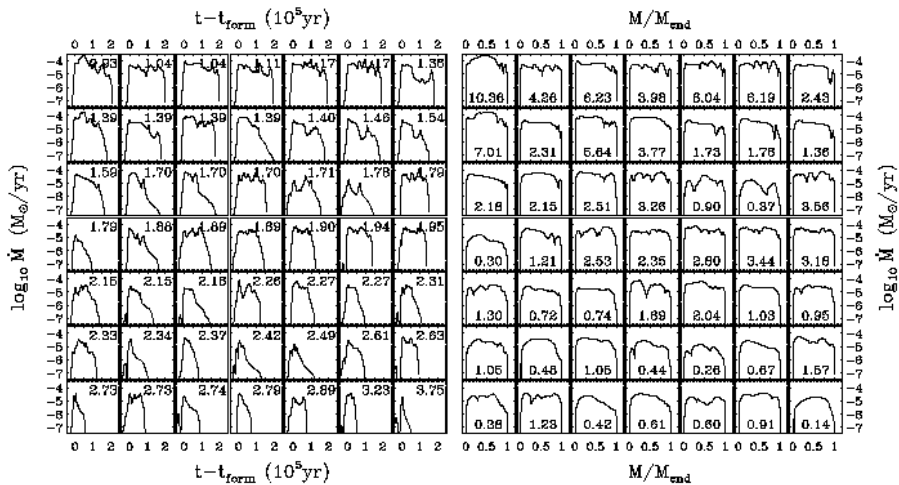}}
\end{picture}
\caption{\label{fig:accretion-rates} Time-varying protostellar mass
  accretion rates in a dense cluster environment. The left panel shows
  accretion rate $\dot{M}$ versus time after formation $t-t_{\rm
    form}$ for 49 randomly selected protostellar cores in a numerical
  model of molecular cloud fragmentation from Klessen \& Burkert
  (2000).  Formation time $t_{\rm form}$ is defined by the first
  occurance of a hydrostatic object in the interior of a collapsing
  gas clump. To link individual accretion histories to the overall
  cluster evolution, $t_{\rm form}$ is indicated in the upper right
  corner of each plot and measures the elapsed time since the start of
  the simulation. The free-fall timescale of the considered molecular
  region is $\tau_{\rm ff} \approx 10^5\,$years.  The right panel
  plots for the same cores $\dot{M}$ as function of the accreted mass
  $M$ with respect to the final mass $M_{\rm end}$, which is indicated
  in the center of each plot. Note that the mass range spans two
  orders of magnitude.  (From Klessen 2001a.)}
\end{figure}

Klessen (2001a) considers the formation of a nascent star cluster for
the case where turbulence is decayed and has left behind random Gaussian
fluctuations in the density structure.  As the system contracts
gravitationally, a dense cluster of protostellar cores builds up on a
timescale of about two to three free-fall times. The protostellar
accretion rates in this environment are strongly time variable, as
illustrated in Figure \ref{fig:accretion-rates}, which is a direct result of
the mutual dynamical interaction and competition between protostellar
cores. While gas clumps collapse to build up protostars, they may
merge as they follow the flow pattern towards the cluster potential
minimum. The timescales for both processes are comparable. The density
and velocity structure of merged gas clumps generally differs
significantly from their progenitor clumps, and the predictions for
isolated cores are no longer valid.  More importantly, these new
larger clumps contain multiple protostars, which subsequently
compete with each other for the accretion from a common gas reservoir.
The most massive protostar in a clump is hereby able to accrete more matter
than its competitors (also Bonnell \etal\ 1997, Klessen \& Burkert
2000, Bonnell \etal\ 2001). Its accretion rate is enhanced through
the clump merger, whereas the accretion rate of low-mass cores
typically decreases.  Temporary accretion peaks in the wake of clump
mergers are visible in abundance in Figure \ref{fig:accretion-rates}.
Furthermore, the small aggregates of cores that build up are
dynamically unstable and low-mass cores may be ejected. As they leave
the high-density environment, accretion terminates and their final
mass is reached.
 
The typical density profiles of gas clumps that give birth to
protostars exhibit a flat inner core, followed by a density fall-off
$\rho \propto r^{-2}$, and are truncated at some finite radius (see
Figure 13 in Klessen \& Burkert 2000), which in the dense centers of
clusters often is due to tidal interaction with neighboring cores.  As
result, a short-lived initial phase of strong accretion occurs when
the flat inner part of the pre-stellar clump collapses.  This
corresponds to the class 0 phase of protostellar evolution (Andr{\'e}
\etal\ 2000). If these cores were to remain isolated and unperturbed,
the mass growth rate would gradually decline in time as the outer
envelope accretes onto the center. This is the class I phase.  Once
the truncation radius is reached, accretion fades and the object
enters the class II phase.  This behavior is expected from analytical
models (e.g.\ Henriksen \etal\ 1997) and agrees with other numerical
studies (e.g.\ Foster \& Chevalier 1993).  However, collapse does not
start from rest for the density fluctuations considered here, and the
accretion rates exceed the theoretically predicted values even for the
most isolated objects in the simulation.

The most massive protostars begin to form first and continue to
accrete at high rate throughout the entire cluster evolution. As the
most massive gas clumps tend to have the largest density contrast,
they are the first to collapse and constitute the center of the
nascent cluster.  These protostars are fed at high rate and gain mass
very quickly. As their parental clumps merge with others, more gas is
fed into their `sphere of influence'. They are able to maintain or
even increase the accretion rate when competing with lower-mass
objects (e.g.\ core 1 and 8 in Figure \ref{fig:accretion-rates}).
Low-mass stars, on average, tend to form somewhat later in the
dynamical evolution of the system (as indicated by the absolute
formation times in Figure \ref{fig:accretion-rates}), and typically
have only short periods of high accretion.

As high-mass stars are associated with large core masses,
while low-mass stars come from low-mass cores, the stellar population
in clusters is predicted to be mass segregated right from the
beginning. High-mass stars form in the center, lower-mass stars tend
to form towards the cluster outskirts. This is in agreement with
recent observational findings for the cluster NGC330 in the Small
Magellanic Cloud (Sirianni \etal\ 2002). Dynamical effects during the
embedded phase of star cluster evolution will enhance this initial
segregation even further.

\section{Mass Spectra from Turbulent Fragmentation}
\label{sec:IMF}
\begin{figure}[bthp]
 \unitlength1.0cm
 \begin{picture}(11.0,8.8)
 \put( 0.0,-0.2){\epsfxsize=11.7cm\epsfbox{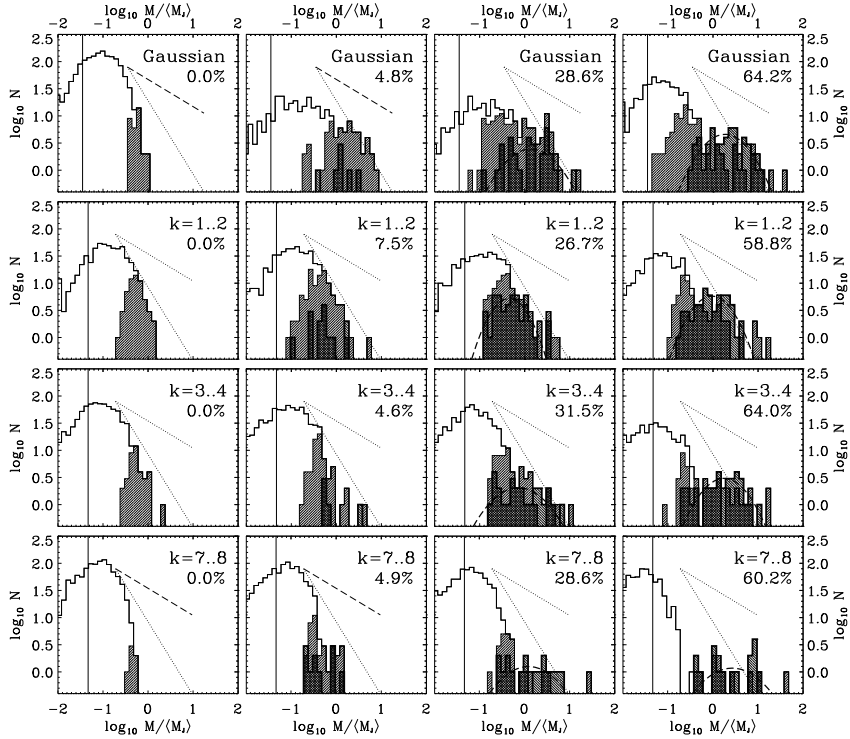}}
 \end{picture}
 \caption{\label{fig:massspectra} Mass spectra of dense collapsed
   cores (hatched thick-lined histograms), of gas clumps (thin lines),
   and of the subset of Jeans unstable clumps (thin lines, hatched
   distribution) for four turbulence models. Gaussian density
   perturbations without turbulence leads to global collapse, while
   three models with turbulence are nominally supported on driven on
   long, intermediate or short scales, respectively, as indicated by
   the driving wavenumbers $k$.  Masses are binned logarithmically and
   normalized to the average Jeans mass $\langle m_{\rm J}\rangle$.
   The left column gives the initial state of the system, the second
   column shows the mass spectra when $m_{\rm *} \approx 5$\% of the
   mass is accreted onto dense cores, the third column shows $m_{\rm
     *} \approx 30$\%, and the last one $m_{\rm *} \approx 60$\%. For
   comparison with power-law spectra ($dN/dm \propto m^{\nu}$), a
   slope $\alpha = -1.5$ typical for the observed clump mass
   distribution, and the Salpeter (1955) slope $\alpha=-2.33$ for the
   IMF, are indicated by the dotted lines.  The vertical line shows
   the resolution limit of the numerical model.  In columns 3 and 4,
   the long dashed curve shows the best log-normal fit. (From Klessen
   2001b.)  }
\end{figure}

As discussed before, a full understanding of turbulent molecular cloud
fragmentation should in principle allow for a prediction of the
distribution of stellar masses (e.g.\ Larson 1981, Fleck 1982,
Elmegreen 1993, Padoan 1995, Padoan \& Nordlund). However, a complete
theory of compressible interstellar turbulence is still out of reach,
and we have to resort to numerical modeling instead to make some
progress.  To illustrate this point we examine the mass spectra of gas
clumps and collapsed cores from models of self-gravitating,
isothermal, supersonic turbulence driven with different wavelengths
(Klessen 2001b).  In the absence of magnetic fields and more accurate
equations of state, these models can only be illustrative, not
definitive, but nevertheless they offer insight into the processes
acting to form the initial stellar mass function (IMF; for a review
see Kroupa 2002).  Figure~\ref{fig:massspectra} plots for four
different models the mass distribution of gas clumps, of the subset of
gravitationally unstable clumps, and of collapsed cores, at four
different evolutionary phases.  In the initial phase, before local
collapse begins to occur, the clump mass spectrum is not well
described by a single power law.  During subsequent evolution, as
clumps merge and grow bigger, the mass spectrum extends towards larger
masses, approaching a power law with slope $\alpha \approx -1.5$.
Local collapse sets in and results in the formation of dense cores
most quickly in the freely collapsing model.  The influence of gravity
on the clump mass distribution weakens when turbulence dominates over
gravitational contraction on the global scale, as in the other three
models. The more the turbulent energy dominates over gravity, the more
the spectrum resembles the initial case of pure hydrodynamic
turbulence. This suggests that the clump mass spectrum in molecular
clouds will be shallower in regions where gravity dominates over
turbulent energy. This may explain the observed range of slopes for
the clump mass spectrum in different molecular cloud regions (e.g.\ 
Kramer \etal\ 1998).

Like the distribution of Jeans-unstable clumps, the mass spectrum of
dense protostellar cores resembles a log-normal in the model without
turbulent support and in the one with long-wavelength turbulent
driving, with a peak at roughly the average thermal Jeans mass
$\langle m_{\rm J}\rangle$ of the system.  However, models supported
at shorter wavelength have mass spectra much flatter than observed,
suggesting that clump merging and competitive accretion are important
factors leading to a log-normal mass spectrum.  The protostellar
clusters discussed here only contain between 50 and 100 cores. This
allows for comparison with the IMF only around the characteristic mass
scale, typically about 1$\,$M$_{\odot}$, since the numbers are too
small to study the very low- and high-mass end of the distribution.
Focusing on low-mass star formation, however, Bate, Bonnell, \& Bromm
(2002) demonstrate that brown dwarfs are a natural and frequent
outcome of turbulent fragmentation. In this model, brown dwarfs form
when dense molecular gas fragments into unstable multiple systems that
eject their smallest members from the dense gas before they have been
able to accrete to stellar masses.  Numerical models with sufficient
dynamic range to treat the full range of stellar masses remain yet to
be done.

\section{Scales of Interstellar Turbulence}
\label{sec:scales}

Turbulence has self-similar properties only within a certain range of
scales. The upper end typically is associated with the global extent
of the system or with the scale at which energy is
inserted. The lower scale is the energy dissipation scale of the
system where turbulent kinetic energy is converted into random motion,
into heat. In purely hydrodynamic systems this is the scale where
molecular viscosity becomes important.  

In interstellar clouds this situation may be different. It was first
shown by Zweibel \& Josafatsson (1983) that ambipolar diffusion would
be the most important dissipation mechanism in typical molecular
clouds with very low ionization fractions $x = \rho_i/\rho_n$, where
$\rho_i$ is the density of ions, $\rho_n$ is the density of neutrals,
and $\rho = \rho_i + \rho_n$.  The ambipolar diffusion strength is
defined as 
\begin{equation}
\lambda_{\rm AD} = v_{\rm A}^2 / \nu_{ni},
\end{equation}
where $v_{\rm A}^2 = B^2/4\pi\rho_n$ approximates the effective Alfv\'en
speed for the coupled neutrals and ions if $\rho_n \gg \rho_i$, and
$\nu_{ni} = \gamma \rho_i$ is the rate at which each neutral is hit by
ions.  The coupling constant depends on the cross-section for
ion-neutral interaction, and for typical molecular cloud conditions
has a value of $\gamma \approx 9.2 \times
10^{13}$~cm$^3\,$s$^{-1}$g$^{-1}$ (e.g.\ Smith \& Mac Low 1997).
Zweibel \& Brandenburg (1997) define an ambipolar diffusion Reynolds
number as
\begin{equation}
R_{\rm AD} = \tilde{L}\tilde{V} / \lambda_{\rm AD} = {\cal M}_{\rm A} \tilde{L} \nu_{ni}/v_{\rm A},
\end{equation}
which must fall below unity for ambipolar diffusion to be important,
where $\tilde{L}$ and $\tilde{V}$ are the characteristic length and
velocity scales, and ${\cal M}_{\rm A} = \tilde{V}/v_{\rm A}$ is the characteristic
Alfv\'en Mach number. In our situation we again can take the rms velocity as
typical value for $\tilde{V}$.
By setting $R_{\rm AD} = 1$, we can  derive a critical lengthscale
below which ambipolar diffusion is important
\begin{equation} \label{lcrit1}
\tilde{L}_{cr} = \frac{v_{\rm A}}{{\cal M}_{\rm A} \nu_{ni}} 
         \approx (0.041 \mbox{ pc})\left(\frac{B}{10\,\mu{\rm G}}\right) {\cal M_{\rm
         A}}^{-1} \left(\frac{x}{10^{-6}}\right)^{-1}
         \left(\frac{n_n}{10^3\,{\rm cm}^{-3}}\right)^{-3/2},
\end{equation}
with the magnetic field strength $B$, the ionization fraction $x$, the
neutral number density $n_n$, and where we have taken $\rho_n = \mu
n_n$, with $\mu = 2.36\,m_{\rm H}$.  This is consistent with typical
sizes of protostellar cores (e.g.\ Bacmann \etal\ 2000), if we assume
that ionization and magnetic field both depend on the density of the
region and follow the empirical laws $n_i = 3 \times 10^{-3}\,{\rm
  cm}^{-3}\,(n_n / 10^5\,{\rm cm}^{-3})^{1/2}$ (e.g.\ Mouschovias
1991) and $B \approx 30\,\mu{\rm G}\, (n_n/10^3\,{\rm
  cm}^{-3})^{1/2}$ (e.g.\ Crutcher 1999).

On large scales, an maximum upper limit to the turbulent cascade in
the Milky Way is given by the extent and thickness of the Galactic
disk. This is indeed the true upper scale, if molecular clouds are
created by converging large-scale flows generated by the collective
influence of recurring supernovae explosions. For individual molecular
clouds this means, that turbulent energy is fed in at scales beyond
the size of the considered cloud. The bulk of its turbulent energy
content is then generated in the event of cloud assembly, which then
dissipates rapidly resulting in short cloud life times. The same
compressional motions that created the cloud in the first place,
however, may also act as continuing source of kinetic energy during
some initial period (e.g.\ Walder \& Folini 2000, Hartmann,
Ballesteros-Paredes, \& Bergin 2001), thus extending the overall cloud
lifetime to a few crossing times.  This energy cascades down to supply
turbulence on smaller scales within the cloud. This picture of
molecular cloud turbulence being driven by large-scale, external
sources is strongly support by analysis of velocity structure which is
always is dominated by the large-scale modes in all clouds observed
(Ossenkopf \& Mac Low 2002).

\section{Efficiency of Star Formation}
\label{sec:eff}

The {\em global} star formation efficiency in normal molecular clouds
is usually estimated to be of the order of a few per cent. Their life
times may be on the order of a few crossing times, i.e.\ a few $10^6$
years (Ballesteros-Paredes \etal\ 1999a, Fukui \etal\ 1999,
Elmegreen 2000).  In this case nearly all models of interstellar
turbulence discussed above are consistent with the observed overall
efficiencies. If molecular clouds survive for several tens of their
free-fall time $\tau_{\rm ff}$ (i.e.\ a few $10^7$ years as proposed
by Blitz \& Shu 1980), turbulence models are more strongly
constrained. However, even in this case models with parameters
reasonable for Galactic molecular clouds can maintain global
efficiencies below $M_* = 5$\% for $~10\,\tau_{\rm ff}$ (Klessen
\etal\ 2000).  Furthermore, it needs to be noted that the {\em local}
star formation efficiency in molecular clouds can reach very high
values. For example, the Trapezium star cluster in Orion is likely to
have formed with an efficiency of about 50\% (Hillenbrand \& Hartmann
1998), compared to a value of 5\% proposed for Taurus-Aurigae.

\section{Termination of Local Star Formation}
\label{sec:termination}
It remains quite unclear what terminates stellar birth on scales of
individual star forming regions, and even whether these processes are
the primary factor determining the overall efficiency of star
formation in a molecular cloud.  Three main possibilities exist.
First, feedback from the stars themselves in the form of ionizing
radiation and stellar outflows may heat and stir surrounding gas up
sufficiently to prevent further collapse and accretion.  Second,
accretion might peter out either when all the high density,
gravitationally unstable gas in the region has been accreted in
individual stars, or after a more dynamical period of competitive
accretion, leaving any remaining gas to be dispersed by the background
turbulent flow.  Third, background flows may sweep through, destroying
the cloud, perhaps in the same way that it was created in the first
place. Most likely the astrophysical truth lies in some combination of
all three possibilities.

If a stellar cluster formed in a molecular cloud contains OB stars,
then the radiation field and stellar wind from these high-mass stars
strongly influence the surrounding cloud material. The UV flux ionizes
gas out beyond the local star forming region. Ionization heats the
gas, raising its Jeans mass, and possibly preventing further
protostellar mass growth or new star formation.  The termination of
accretion by stellar feedback has been suggested at least since the
calculations of ionization by Oort \& Spitzer (1955). Whitworth (1979)
and Yorke \etal\ (1989) computed the destructive effects of individual
blister H{\sc ii} regions on molecular clouds, while  Franco \etal\ (1994) and
Diaz-Miller \etal\ (1998) concluded that indeed the ionization from
massive stars may limit the overall star forming capacity of molecular
clouds to about 5\%.  Matzner (2002) analytically modeled the effects
of ionization on molecular clouds, concluding as well that turbulence
driven by H{\sc ii} regions could support and eventually destroy
molecular clouds.  The key question facing these models is whether
H{\sc ii} region expansion couples efficiently to clumpy,
inhomogeneous molecular clouds, a question probably best addressed
with numerical simulations.

Bipolar outflows are a different manifestation of protostellar
feedback, and may also strongly modify the properties of star forming
regions (Norman \& Silk 1980, Adams \& Fatuzzo 1996).  Recently
Matzner \& McKee (2000) modeled the ability of bipolar outflows to
terminate low-mass star formation, finding that they can limit star
formation efficiencies to 30--50\%, although they are ineffective in
more massive regions.  How important these processes are compared to
simple exhaustion of available reservoirs of dense gas (Klessen \etal\ 
2000) remains an important question.

The models relying on exhaustion of the reservoir of dense gas argue
that only dense gas will actually collapse, and that only a small
fraction of the total available gas reaches sufficiently high
densities, due to cooling (Schaye 2002), gravitational collapse and
turbulent triggering (Elmegreen 2002), or both (Wada, Meurer, \&
Norman 2002). This of course pushes the question of local star
formation efficiency up to larger scales, which may indeed be the
correct place to ask it.

Other models focus on competitive accretion in local star formation,
showing that the distribution of masses in a single group or cluster
can be well explained by assuming that star formation is fairly
efficient in the dense core, but that stars that randomly start out
slightly heavier tend to fall towards the center of the core and
accrete disproportionately more gas (Bonnell \etal\ 1997, 2001).
These models have recently been called into question by the
observation that the stars in lower density young groups in Serpens
simply have not had the time to engage in competitive accretion, but
still have a normal IMF (Olmi \& Testi 2002).

Finally, star formation in dense clouds created by turbulent flows may
be terminated by the same flows that created them.  Ballesteros-Paredes
\etal\ (1999a) suggested that the coordination of star formation over
large molecular clouds, and the lack of post-T Tauri stars with ages
greater than about 10$\,$Myr tightly associated with those clouds,
could be explained by their formation in a larger-scale turbulent
flow.  Hartmann \etal\ (2001) make the
detailed argument that these flows may disrupt the clouds after a
relatively short time, limiting their star formation efficiency that
way.  It can be argued that field supernovae are the most likely
driver for this background turbulence in spiral galaxies like the
Milky Way (Mac~Low \& Klessen 2003).

\section{Summary: The Control of Star Formation by Supersonic Turbulence}
\label{sec:summary}

In this review we have proposed that star formation is regulated by
interstellar turbulence and its interplay with gravity. We have
discussed that this new approach can explain the same observations
successfully described by the so called ``standard theory'', while
also addressing (and resolving!) its inconsistencies with other observed
properties of Galactic star forming regions.

The key point to this new understanding of star formation in Galactic
molecular clouds lies in the properties of interstellar turbulence.
Turbulence is observed in the interstellar medium almost ubiquitously
and is typically supersonic as well as super-Alfv{\'e}nic. It is
energetic enough to counterbalance gravity on global scales, but at
the same time it may provoke local collapse on small scales. This
apparent paradox can be resolved when considering that supersonic
turbulence establishes a complex network of interacting shocks, where
converging flows generate regions of high density. This density
enhancement can be sufficiently large for gravitational instability to
set in.  The same random flow that creates density enhancements,
however, may disperse them again.  For local collapse to result in
stellar birth, it must progress sufficiently fast for the region to
`decouple' from the flow.  Typical collapse timescales are hereby of
the same order as the lifetimes of shock-generated density
fluctuations in the turbulent gas. This makes the outcome highly
unpredictable. As stars are born through a sequence of stochastic
events, any theory of star formation is in essence a statistical one
with quantitative predictions only possible for an ensemble of stars.

In the new picture, the efficiency of protostellar core formation, the
growth rates and final masses of the protostars, essentially all
properties of nascent star clusters depend on the intricate interplay
between gravity on the one hand side and the turbulent velocity field
in the cloud on the other. The star formation rate is regulated not
just  at the scale of individual star-forming cores through
ambipolar diffusion balancing magnetostatic support, but rather at all
scales (Elmegreen 2002), via the dynamical processes that determine
whe\-ther regions of gas become unstable to prompt gravitational
collapse. The presence of magnetic fields does not alter that picture
significantly, as long as they are too weak for magnetostatic support,
which is indicated by observations (Crutcher 1999, Bourke \etal\
2001). In particular, magnetic fields cannot prevent the decay of interstellar
turbulence, which in turn needs to be continuously driven or else
stars form quickly and with high efficiency

Inefficient, isolated star formation will occur in regions which are
supported by turbulence carrying most of its energy on very small
scales. This typically requires an unrealistically large number of
driving sources and appears at odds with the measured velocity
structure in molecular clouds which in almost all cases is dominated
by large-scale modes. The dominant pathway to star formation therefore
seems to involve cloud regions large enough to give birth to
aggregates or clusters of stars. This is backed up by careful stellar
population analysis indicating that most stars in the Milky Way formed
in open clusters with a few hundred member stars (Adams \& Myers
2001).

Clusters of stars build up in molecular cloud regions where
self-gravity overwhelms turbulence, either because such regions are
compressed by a large-scale shock, or because interstellar turbulence
is not replenished and decays on short timescales.  Then, many
gas clumps become gravitationally unstable synchronously and start to
collapse. If the number density is high, collapsing gas clumps may
merge to produce new clumps which now contain multiple protostars.
Mutual dynamical interactions become common, with close encounters 
drastically altering the protostellar trajectories, thus changing the
mass accretion rates. This has important consequences for the IMF.  Already
in their infancy, i.e.\ already in the deeply embedded phase, very
dense stellar clusters are expected to be strongly influenced by
collisional dynamics.

\subsection*{Acknowledgments}
This review would not have been possible without long-term
collaboration and exchange of ideas with M.-M.\ Mac~Low. Special
thanks also to P.\ Bodenheimer and D.\ Lin for many vivid scientific
discussions and for their warm hospitality at UC Santa Cruz; and
thanks to J.\ Ballesteros-Paredes, F.\ Heitsch, P.\ Kroupa, E.\ 
V\'azquez-Semadeni, and H.\ Zinnecker.

I want to express my gratitudes to the members of the Astronomische
Gesellschaft for awarding the Ludwig Biermann Preis to me; in
particular, I want to thank G.\ Hensler and A.\ Burkert in this context.  I
furthermore acknowledge support by the Emmy Noether Program of the
Deutsche Forschungsgemeinschaft (DFG: KL1358/1).

\subsection*{References}

\def\and{{and }}

{\small

\bref Adams, F.\ C., \and M.\ Fatuzzo, 1996, \apj, {\bf 464}, 256

\bref Adams, F.\ C., \and P.\ C.\ Myers, 2001, \apj, {\bf 553}, 744

\bref Andr{\'e}, P., D.\ Ward-Thompson, \and M.\ Barsony, 2000, in
{\em Protostars and Planets IV}, edited by V.\ Mannings, A.\ P.\ Boss,
\and S.\ S.\ Russell (University of Arizona Press, Tucson), p.\ 59

\bref Bacmann, A., P.\ Andr{\'e}, J.\ -.L\ Puget, A.\ Abergel,
S.\  Bontemps, \and D.\ Ward-Thompson,  2000,  \aap,  {\bf 361}, 555  

\bref  Ballesteros-Paredes, J., L.\ Hartmann, \and E.\ V{\'
a}zquez-Semadeni, 1999a, \apj, {\bf 527}, 285 

\bref Ballesteros-Paredes, J., E.\ V{\'a}zquez-Semadeni, \and J.\ Scalo,
1999b,  \apj,  {\bf 515}, 286  

\bref Balsara, D. S., R.\ M.\ Crutcher, \and A.\ Pouquet, 2001, \apj, {\bf 557}, 451

\bref Bate, M.\ R., I.\ A.\ Bonnell, \and V.\ Bromm, 2002, \mnras, {\bf 332}, L65

\bref Benz, W., 1990, in {\em The Numerical Modelling of Nonlinear
Stellar Pulsations}, edited by J. R. Buchler (Kluwer, Dordrecht), 269

\bref Bergin, E.\ A., \and W.\ D.\ Langer,  1997,  \apj,  {\bf 486}, 316 

\bref Biskamp, D., \and W.-C.\ M\"uller, 2000, \ppl, {\bf 7}, 4889

\bref Blitz, L.,\ \and F.\ H.\ Shu, 1980, \apj, {\bf 238}, 148

\bref Boldyrev, S., 2002, \apj, {\bf 569}, 841

\bref Boldyrev, S., \AA.\ Nordlund, \and P.\ Padoan, 2002, \apj, {\bf 573}, 678 


\bref Bonazzola, S., E.\ Falgarone, J.\ Heyvaerts, M.\  Perault, \and J.\ L.\ Puget, 1987, \aap, {\bf 172}, 293

\bref Bonnell, I.\ A., M.\ R.\ Bate, C.\ J.\ Clarke, \and J.\ E.\  Pringle,  1997,  \mnras,  {\bf 285}, 201 

\bref Bonnell, I.\ A., M.\ R.\ Bate, C.\ J.\ Clarke, \and J.\ E.\ Pringle, 2001, \mnras, {\bf 323}, 785

\bref Bourke, T.\ L., P.\ C.\ Myers, G.\ Robinson, \and A.\ R.\ Hyland,
2001,  \apj,  {\bf 554}, 916  

\bref Chandrasekhar, S., 1949, \apj, {\bf 110}, 329

\bref Crutcher, R.\ M.,  1999,  \apj,  {\bf 520}, 706 

\bref Desch, S.\ J.\ \and T.\ C.\ Mouschovias,  2001,  \apj,  {\bf 550},  314  

\bref Dewar, R.\ L., 1970, \pfl, {\bf 13}, 2710

\bref Diaz-Miller, R.~I., J.~Franco, \and S.~N.~Shore, 1998, \apj,
{\bf 501}, 192.

\bref Efremov, Y.\ N., \and B.\ G.\ Elmegreen,  1998,  \mnras,
{\bf 299},  588  

\bref Elmegreen, B.\ G., 1991,  in {\em NATO ASIC Proc.\ 342: The
Physics of Star Formation and Early Stellar Evolution},  edited by C.\
J.\ Lada \and N.\ D.\ Kylafis (Kluwer Academic Publishers), p.\ 35

\bref Elmegreen, B.\ G., 1993, \apj, {\bf 419}, L29

\bref Elmegreen, B.\ G., 2000, \mnras, {\bf 311}, L5

\bref Elmegreen, B.\ G., 2002, \apj, {\bf 577}, 206

\bref Fiege, J.\ D., \and R.\ E.\ Pudritz, 2000, \mnras, {\bf 311}, 85

\bref Field, G.\ B., D.\ W.\ Goldsmith, \and H.\ J.\ Habing, 1969,
\apj, {\bf 155}, L49 

\bref Fleck, R.\ C., 1982, \mnras, {\bf 201}, 551

\bref Foster, P.\ N., \and R.\ A.\ Chevalier, 1993, \apj, {\bf 416}, 303

\bref Franco, J., S.~N.~Shore, \and G.~Tenorio-Tagle, 1994, \apj,
{\bf 436}, 795

\bref Fukui, Y.\ {\em et al.}, 1999, \pasj, {\bf 51}, 745 

\bref Gammie, C.\ F., \and E.\ C.\ Ostriker, 1996, \apj, {\bf 466}, 814

\bref Goldreich, P.\ \and S.\ Sridhar, 1995, \apj, {\bf 438}, 763

\bref Goldreich, P.\ \and S.\ Sridhar, 1997, \apj, {\bf 485}, 680

\bref Hartmann, L., 2001, \aj, {\bf 121}, 1030

\bref Hartmann, L., J.\ Ballesteros-Paredes, \and E.\ A.\ Bergin, 2001,
\apj, {\bf 562}, 852  

\bref Heitsch, F., M.\ Mac Low, \and R.\ S.\ Klessen,  2001a,
  \apj,   {\bf 547}, 280 

\bref Heitsch, F., E.\ G.\ Zweibel, M.-M.\ Mac Low, P.\ Li, and
M.\ L.\ Norman, 2001b, \apj, {\bf 561}, 800

\bref Herbig, G, 2002, in {\em Saas Fee Advanced Course 29: Physics of
  Star Formation in Galaxies}, edited by F.\ Palla, H.\ Zinnecker, A.\ 
Maeder, \and G.\ Meynet (Springer, Berlin, Heidelberg), p.\ 1


\bref Hillenbrand, L.\ A., \and L.\ W.\ Hartmann, 1998, \apj, {\bf 492}, 540

\bref Hendriksen, R.\ N., P.\ Andr{\'e}, \and S.\ Bontemps, 1997, \aap, {\bf 323}, 549

\bref Jeans, J.\ H., 1902, \pta, {\bf 199}, 1

\bref Klein, R.\ I., C.\ F.\ McKee, \and P.\ Colella, 1994, \apj,
{\bf 420}, 213

\bref Klessen, R. S., 2000, \apj, {\bf 535}, 869

\bref Klessen, R.\ S., \and A.\ Burkert,  2000,  \apjs,  {\bf 128}, 287 

\bref Klessen, R.\ S., \and A.\ Burkert,  2001,  \apj,  {\bf 549}, 386 

\bref Klessen, R.\ S., A.\ Burkert, \and M.\ R.\ Bate, 1998, \apj, {\bf 501}, L205 

\bref Klessen, R.\ S., F.\ Heitsch, \and M.-M.\ Mac Low,  2000,
\apj,   {\bf 535}, 887  

\bref Kolmogorov, A.\ N., 1941, \dansssr, {\bf 30}, 301 (reprinted in
{\em Proc.\ R.\ Soc.\ Lond.\ A}, {\bf 434}, 9-13 [1991])

\bref Kramer, C., J.\ {Stutzki}, R.\ {Rohrig},  U.\ {Corneliussen}, 1998, \aap, {\bf 329}, 249 

\bref Krebs, J., \and W.\ Hillebrandt, 1983, \aap, {\bf 128}, 411

\bref Kroupa, P., 2002, {\em Science}  {\bf 295}, 82 

\bref Lada, E.\ A., 1992, \apj, {\bf 393}, L25

\bref Larson, R.\ B.,  1981, \mnras,  {\bf 194}, 809

\bref Larson, R.\ B.,  1995, \mnras,  {\bf 272}, 213

\bref Leisawitz, D., F. N. Bash, \and P. Thaddeus, 1989, \apjs, {\bf 70},
731 

\bref L\'eorat, J., T.\ Passot, \and A.\ Pouquet, 1990, \mnras, {\bf 243},
293 

\bref Lesieur, M., 1997, {\em Turbulence in Fluids}, 3rd ed.\
(Kluwer, Dordrecht), p.\ 245 

\bref Lithwick, Y., \and P.\ Goldreich, 2001, \apj, {\bf 562}, 279

\bref Mac Low, M.-M., 1999, \apj, {\bf 524}, 169

\bref Mac~Low, M.-M., Klessen, R.\ S., 2003, \rmp, in press (astro-ph/0301093)

\bref Mac Low, M.-M., R.\ S.\ Klessen, A.\ Burkert, \and M.\ D.\
Smith, 1998, \prl, {\bf 80}, 2754 

\bref Mac Low, M.-M.,, C.\ F.\ McKee, R.\ I.\ Klein, J.\ M.\
Stone, \and M.\ L.\ Norman, 1994, \apj, {\bf 433}, 757

\bref Matzner, C.\ D., 2002, \apj, {\bf 566}, 302
  
\bref Matzner, C.\ D., \and C.\ F.\ McKee, 2000, \apj, {\bf 545},
  364

\bref McKee,  C.\ F., 1999, in {\em NATO ASIC Proc.\ 540: The
Origin of Stars  and Planetary Systems},  edited by C.\ J.\ Lada and
N.\ D.\ Kylafis (Kluwer Academic Publishers), p.\ 29 

\bref McKee, C.\ F., \and J.\ P.\ Ostriker, 1977, \apj, {\bf 218},
148

\bref McKee, C.\ F., \and E.\ G.\ Zweibel, 1995, \apj, {\bf 440},
686

\bref Mizuno, A., T.\ Onishi, Y.\ Yonekura, T.\ Nagahama, H.\
Ogawa, \and Y.\ Fukui, 1995, \apj, {\bf 445}, L161

\bref Monaghan, J. J., 1992, \araa, {\bf 30}, 543


\bref Mouschovias, T.\ C., 1991, in {\em The Physics of Star
Formation and Early Stellar Evolution}, edited by C.\ J.\ Lada \and N.\ D.\
Kylafis (Kluwer, Dordrecht), p.\ 449

\bref Mouschovias, T.\ C., \and L.\ Spitzer, Jr., 1976, \apj, {\bf 210}, 326 

\bref M\"uller, W.-C., \and D.\ Biskamp, 2000, \prl, {\bf 84}, 475

\bref Nakano, T., 1976, \pasj,  {\bf 28}, 355

\bref Nakano, T., 1998, \apj, {\bf 494}, 587

\bref Nakano, T., \and T.\ Nakamura, 1978, \pasj, {\bf 30}, 681

\bref Ng, C.\ S., \and A.\ Bhattacharjee, 1996, \apj, {\bf 465}, 845

\bref Norman, C.\ A., \and A.\ Ferrara, 1996, \apj, {\bf 467}, 280

\bref Norman, C.\ A., \and J.\ Silk, 1980, \apj,  {\bf 239}, 968 

\bref Olmi, L., \and L. Testi, 2002, \aap, {\bf 392}, 1053

\bref Oort, J. H., \& L. Spitzer, Jr., 1955, \apj, {\bf 121}, 6

\bref Ostriker, E.\ C., C.\ F.\ Gammie, \and J.\ M.\ Stone, 1999, \apj, {\bf 513}, 259

\bref Ossenkopf V., \and M.-M.\ Mac\ Low, 2002, \aap,  {\bf 390}, 307 

\bref Padoan, P., 1995, \mnras, {\bf 277}, 377

\bref Padoan, P., \and \AA.\ Nordlund, 1999, \apj, {\bf 526}, 279 

\bref Padoan, P., \and \AA.\ Nordlund, 2002, \apj, {\bf 576}, 870 

\bref Palla, F., \and S.\ W.\ Stahler, 2000, \apj, {\bf 540}, 255

\bref Passot, T., A.\ Pouquet, \and P.\ R.\ Woodward,  1988, \aap, {\bf 197}, 392

\bref Passot, T., E.\ V{\'a}zquez-Semadeni, \and A.\ Pouquet, 1995, \apj, {\bf 455}, 536

\bref Porter, D.\ H., A.\ Pouquet, \and P.\ R.\ Woodward, 1994, \pfl, {\bf
6}, 2133 


\bref Salpeter, E.\ E., 1955, \apj, {\bf 121}, 161

\bref Scalo, J.\ M., E.\ V{\'a}zquez-Semadeni, D.\ Chappell, T.\
Passot,  1998, \apj, {\bf 504}, 835   

\bref Schaye, J., 2002, \apj, submitted (astro-ph/0205125)

\bref She, Z., \and E.\ Leveque, 1994, \prl, {\bf 72}, 336

\bref Shu, F.\ H., 1977, \apj,  {\bf 214}, 488

\bref Shu, F.\ H., F.\ C.\ Adams, \and S.\ Lizano,  1987,  \araa,
 {\bf 25}, 23  

\bref Sirianni, M., A.\ Nota, G.\ De Marchi, C.\ Leitherer, and
M.\ Clampin, 2002, \apj, {\bf 579}, 275

\bref Smith, M.\ D., \and M.-M.\ Mac Low,  1997, \aap, {\bf 326}, 801

\bref Spaans, M., \and J., Silk, 2000, \apj, {\bf 538}, 115

\bref Stone, J. M., \and M. L. Norman, 1992a, \apjs, {\bf 80}, 753

\bref Stone, J. M., \and M. L. Norman, 1992b, \apjs, {\bf 80}, 791

\bref Stone, J.\ M., E.\ C.\ Ostriker, \and C.\ F.\ Gammie, 1998, \apj, {\bf 508}, L99 

\bref Strom, K.\ M., S.\ E.\ Strom,  \and K.\ M.\ Merrill, 1993, \apj, {\bf 412}, 233 

\bref Tafalla, M., D.\ Mardones, P.\ C.\ Myers, P.\ Caselli, R.\
Bachiller,  \and P.\ J.\ Benson,  1998,  \apj,  {\bf 504}, 900  

\bref V{\'a}zquez-Semadeni, E., T.\ Passot, \and A.\ Pouquet,
  1995, \apj, {\bf 441}, 702 

\bref V{\'a}zquez-Semadeni, E., T.\ Passot, \and A.\ Pouquet, 1996 \apj, {\bf 473}, 881


\bref von Weizs\"acker, C.\ F., 1943, \za, {\bf 22}, 319

\bref von Weizs\"acker, C.\ F., 1951, \apj, {\bf 114}, 165

\bref Wada, K., \and C.\ A.\ Norman,  1999,  \apj,  {\bf 516}, L13 

\bref Wada, K., G. Meurer, \and C.\ A.\ Norman,  2002,  \apj,
{\bf 577}, 197  

\bref Walborn, N.\ R., R.\ H.\ Barb{\'a}, W.\ Brandner, M.\ ;.\
Rubio, E.\ K.\ Grebel, \and R.\ G.\ Probst, 1999, \aj, {\bf 117}, 225 

\bref Walder, R.\ \and D.\ Folini, 2000, \apss, {\bf 274}, 343

\bref Whitworth, A. P., 1979, \mnras, {\bf 186}, 59

\bref Whitworth, A.\ P., A.\ S.\ Bhattal, N.\ Francis, \and S.\
J.\ Watkins, 1996, \mnras, {\bf 283}, 1061 

\bref Williams, J.\ P., L.\ Blitz, \and C.\ F.\ McKee, 2000, in
{\em Protostars and Planets IV},  edited by V.\ Mannings,  A.\ P.\
Boss, \and S.\ S.\ Russell (University of Arizona Press, Tucson), p.\
97 

\bref Williams, J.\ P., P.\ C.\ Myers, D.\ J.\ Wilner, \and J.\ di
Francesco,  1999,  \apj,  {\bf 513}, L61  

\bref Wolfire, M.\ G., D.\ Hollenbach, C.\ F.\ McKee, A.\ G.\ G.\ M.\ Tielens,  \and E.\ L.\ O.\ Bakes, 1995, \apj, {\bf 443}, 152

\bref Yorke, H. W., Tenorio-Tagle, G., Bodenheimer, P., and
M. R\'o\.zyczka, 1989, \aap, {\bf 216}, 207  

\bref Zinnecker, H., M.~J.~McCaughrean, \and B.~A.~Wilking. 1993, in
{\em Protostars and Planets III}, edited by E.\ H.\ Levy \and J.\ I.\ 
Lunine (University of Arizona Press, Tucson), p.\ 429

\bref Zweibel, E.\ G., \and A.\ Brandenburg, 1997, \apj, {\bf 478},
563

\bref Zweibel, E.\ G., \and K.\ Josafatsson, 1983, \apj, {\bf 270}, 511

}

\vfill

\end{document}